\documentclass[iop,revtex4]{emulateapj}
\usepackage{lineno}

\usepackage{lscape}

\begin{document}

% for float placement:
\renewcommand{\topfraction}{1.0}
\renewcommand{\bottomfraction}{1.0}
\renewcommand{\textfraction}{0.0}

\shorttitle{Speckle interferometry at SOAR}
\shortauthors{Tokovinin et al.}

\title{Speckle interferometry at SOAR in 2019 }

\author{Andrei Tokovinin}
\affil{Cerro Tololo Inter-American Observatory,\footnote{National Science Foundation's 
 National Optical-Infrared Astronomy Research Laboratory} Casilla 603, La Serena, Chile}
\email{atokovinin@ctio.noao.edu}

\author{Brian D. Mason}
\affil{U.S. Naval Observatory, 3450 Massachusetts Ave., Washington, DC, USA}
\email{brian.d.mason@navy.mil}
\author{Rene A. Mendez}
\affil{Universidad de Chile,  Casilla 36-D, Santiago, Chile}
\email{rmendez@u.uchile.cl}
\author{Edgardo Costa}
\affil{Universidad de Chile,  Casilla 36-D, Santiago, Chile}
\author{Elliott P. Horch\footnote{Adjunct Astronomer, Lowell Observatory} }
\affil{Department of Physics, Southern Connecticut State University, 501 Crescent Street, New Haven, CT 06515, USA}
\email{horche2@southernct.edu}

%\author{Cesar Brice\~no}
%\affil{Cerro Tololo Inter-American Observatory, Casilla 603, La Serena, Chile}
%\email{cbriceno@ctio.noao.edu}

\begin{abstract}
The  results of  speckle  interferometric observations  at  the 4.1  m
Southern Astrophysical  Research Telescope  (SOAR) in 2019  are given,
totaling  2555 measurements  of 1972  resolved pairs  with separations
from 15 mas  (median 0\farcs21) and magnitude difference up  to 6 mag, and
non-resolutions of 684 targets. We  resolved for the first time 90 new
pairs  or  subsystems  in  known  binaries. This  work  continues  our
long-term speckle program. Its main  goal is to monitor orbital motion
of  close binaries,  including members  of high-order  hierarchies and
{\it Hipparcos} pairs in the solar neighborhood. We give a list of 127
orbits computed  using our  latest measurements. Their  quality varies
from excellent (25 orbits of grades 1 and 2) to provisional (47 orbits
of grades 4 and 5).
\end{abstract} 

%---------------------------------------------------------
\section{Introduction}
\label{sec:intro}

We report  here a  large set of  double-star measurements made  at the
4.1 m  Southern  Astrophysical  Research  Telescope  (SOAR)  with  the
speckle camera,  HRCam.  This paper continues the  series published by
\citet[][hereafter   TMH10]{TMH10},  \citet{SAM09},  \citet{Hrt2012a},
\citet{Tok2012a},  \citet{TMH14},  \citet{TMH15}, 
\citet{SAM15}, \citet{SAM17}, and \citet{SAM18}. The aims are outlined in these
papers and briefly recalled below.  The data were taken during
2019. They are presented in the same format as in \citet{SAM18}. 

Section~\ref{sec:obs} reviews all speckle programs executed at SOAR in
2019.  The results are  presented in Section~\ref{sec:res} in the form
of  electronic tables  archived by  the journal.  We also  discuss new
resolutions and provide a large list of new orbital elements.  A short
summary in Section~\ref{sec:sum} closes the paper.

%---------------------------------------------------------
\section{Observations}
\label{sec:obs}

%-------------------------------------------------------------
\subsection{Observing programs}

As in  previous years, HRCam (see Sect.~\ref{sec:inst})
was  used during 2019 to  execute  several observing  programs,  some with  common
(overlapping) targets.  Table~\ref{tab:programs}  gives an overview of
these programs  and indicates which observations are  published in the
present paper. Here is a brief description of these programs.

{\it  Orbits} of resolved  binaries are  of fundamental  importance in
various  areas of astronomy,  e.g. for  direct measurement  of stellar
masses, binary statistics, astrometry, and objects of special interest
such as binaries hosting exo-planets. Observations of tight pairs with
fast motion, mostly nearby  dwarfs, are prioritized at SOAR.  However,
classical  visual binaries  are also  observed at  low  cadence to
improve their orbits. The Sixth  Catalog of Visual Binary Star Orbits,
VB6 \citep{VB6}, contains a substantial fraction of poorly determined,
low-grade orbits based on  inaccurate and/or sparse visual micrometric
measures.  This situation is  slowly improving.   Our work  has added many
orbits to VB6, more are given here in Section~\ref{sec:orbits}.

{\it  Hierarchical  systems}  of   stars  challenge  the  theories  of
binary-star formation.  Better  observational data on their statistics
and   architecture   (orbits,   relative  inclinations)   are   needed
\citep{MSC}.   Many hierarchies  have  been discovered  at SOAR  using
HRCam, and  we are  following their orbital  motion.  This  paper adds
several newly discovered hierarchies and several orbits of subsystems.

{\it Hipparcos binaries} \citep{HIP} within 200\,pc  are monitored with the aim of
determining orbits and  masses for stars in a  wide range of effective
temperatures      and      metallicities,      as     outlined      by
\citet{Horch2015,Horch2017,Horch2019}.   The  southern  part  of  this
sample is addressed at SOAR \citep{Mendez2017}.  This program overlaps
with  the  general work  on  orbits.   

Accurate  parallaxes  of visual  binaries  combined with  good-quality
orbits will  allow accurate  measurements of stellar  masses. However,
the  parallactic and  orbital motions  are coupled.   The  second {\it
  Gaia} data release, DR2 \citep{Gaia}, uses only a linear 5-parameter
astrometric  model  and contains  examples  of  biased parallaxes  and
proper motions  of tight  visual binaries. Including  acceleration and
higher-order  terms  in  the  astrometric solution  will  improve  the
situation, but the ultimate astrometric precision will be reached only
when  the  orbit is  explicitly  included  in  the astrometric  model.
Considering the limited  duration of the {\it  Gaia} mission, ground-based
coverage  is and will  remain essential  for accurate  measurements of
stellar masses.

\begin{deluxetable}{ l l l l  } 
\tabletypesize{\scriptsize}    
\tablecaption{Observing programs executed with HRCam in 2019
\label{tab:programs} }                    
\tablewidth{0pt}     
\tablehead{ \colhead{Program}  &
\colhead{PI}  &  
\colhead{$N$} & 
\colhead{Publ.\tablenotemark{a}} 
}
\startdata
Orbits                & Mason, Tokovinin    & 996 & Yes \\
Hierarchical systems  & Tokovinin           & 188 & Yes \\
Hipparcos binaries    & Mendez, Horch       & 737 & Yes \\
Neglected binaries    & R.~Gould, Tokovinin & 363  & Yes \\
Binaries in Upper Scorpius   & Tokovinin, Brice\~no, & 485 & Pub \\
Nearby K,M dwarfs     & E. Vrijmoet           & 453 & No \\
TESS follow-up        & C. Ziegler           & 785  & Pub \\
Young moving groups   & A. Mann              & 645  & No \\ 
Stars with RV trends  & B. Pantoja           & 48  & No
\enddata
\tablenotetext{a}{This columns indicates whether the results are
  published here (Yes), previously (Pub), or deferred to future
  papers (No). }
\end{deluxetable}

{\it Neglected  binaries} with  small separations from  the Washington
Double Star Catalog, WDS \citep{WDS} are observed with a low priority,
as a ``filler''.  Lists of pairs in need of fresh data are provided by
R.~Gould. A fraction  of these stars are interesting  because they are
presently very  tight, near the  periastron of their orbits.   Some of
these pairs contain additional, previously unknown, components.

{\it Members  of the  Upper Scorpius} association  were surveyed  in a
systematic way  by \citet{Sco} to find multiplicity  fraction and the
distribution of periods and mass  ratios, taking advantage of the high
productivity  of HRCam.   The work  started in  2018. A  total  of 614
targets  were observed during  2018 and  2019 using  approximately two
nights of telescope time.  Several interesting results are reported in
the above paper. Moreover, new close pre-main sequence pairs with fast
orbital  motion  are excellent  candidates  for  measuring masses  and
testing evolutionary models of young stars.

{\it Nearby K and M dwarfs} were observed for E. Vrijmoet. His program
aims at determination  of a large number of orbits  to throw new light
on the statistics of orbital elements. As these stars are nearby, some
have very short orbital periods and displayed a substantial orbital
motion during 2019. 

{\it TESS  follow-up} was one of the major observing programs during
2019. Its first results are published by \citet{TESS}, but
observations continued since the submission of this paper, and the number
of surveyed TESS objects of interests has almost doubled. 

If observations  of a given  star were requested by  several programs,
they are  published here even if the other program still  continues.  We
also publish  measurements of  previously known pairs  resolved during
surveys, for example in the TESS follow-up.

%-------------------------------------------------------------
\subsection{Instrument and observing procedure}
\label{sec:inst}

The   observations  reported   here  were   obtained  with   the  {\it
  high-resolution camera} (HRCam) -- a fast imager designed to work at
the 4.1 m SOAR telescope \citep{HRCAM}.  The camera was mounted on the
SOAR Adaptive Module \citep[SAM,][]{SAM}.  The laser guide star of SAM
was not  used, the deformable  mirror of SAM was  passively flattened,
and the images are seeing-limited. {\bf However, the atmospheric dispersion corrector (ADC) inside SAM was critical for getting good-quality data.} 
In most observing runs, the median
image  size   was  $\sim$0\farcs6.      The transmission  curves of
HRCam   filters  are   given  in   the   instrument  manual.\footnote{
  \url{http://www.ctio.noao.edu/soar/sites/default/files/SAM/\-archive/hrcaminst.pdf}}
We used  mostly the near-infrared $I$ filter  (824/170\,nm)  and
the Str\"omgren $y$ filter (543/22\,nm);  two measures were made in the
$R$ filter (596/121\,nm) {\bf and two in the H$\alpha$ filter (657.3/5\,nm).}

For  each observing  run, a  selection  of suitable  targets from  all
programs  was  made.   It  contains accurate  coordinates  and  proper
motions  (PMs) to  allow for  precise pointing  of the  telescope. The
slews  are commanded  from the  custom  observing tool  that helps  to
maximize the observing  efficiency. When the slew angle  is small, the
next object  is acquired  almost immediately.  Most  observations were
taken in the narrow 3\arcsec  ~field with the 200$\times$200 region of
interest (ROI), without binning, in the $I$ filter; the $y$ filter was
used  mostly for  brighter and/or  closer pairs.   The pixel  scale is
0\farcs01575 and the  exposure time is normally 24\,ms  (it is limited
by  the camera readout  speed).  Pairs  wider than  $\sim$1\farcs4 are
observed in a  400$\times$400 ROI, and the widest  pairs are sometimes
recorded  with the  full  field  of 1024  pixels  (16\arcsec) and  
2$\times$2 binning. However, the  speckle contrast drops very strongly
at separations  above 3\arcsec, substantially reducing  the quality of
measures of  wide pairs.  Binning  is used mostly  for the fainter
targets; it does not result in the loss of resolution in the $I$ band,
which  ranges  from 40  to  45 mas,  depending  on  the magnitude  and
conditions.   Bright stars  can  be resolved  and  measured below  the
formal diffraction limit by fitting  a model to the power spectrum and
using observations of point  sources as reference.  The resolution and
contrast  limits  of HRCam  are  further  discussed  in TMH10  and  in
previous papers of this series. For each target, two data cubes of 400
frames are normally recorded and processed independently. This ensures
reliability of results despite occasional problems like  cosmic
ray spikes or telescope vibration.

The first  observations reported here  were obtained in  2019 January,
and the  last in 2019 December,  in 9 observing runs.   HRCam was used
during  scheduled observing  time, but  also in  parts  of engineering
nights available from other work.  The total number of observations in
2019 (including reference stars) is 5964; the vast majority (5242) are
made in the  $I$ filter, while for bright and close  pairs we used the
$y$ filter (714  observations).  The full set of  the 2019 data counts
3199 measurements  of 2774 resolved  pairs, mostly (but  not entirely)
published  here. Almost  all  targets are  brighter  than $I=12$  mag,
although several fainter pairs were measured under very good seeing.

%-------------------------------------------------------------
\subsection{Data processing and calibration}
\label{sec:dat}

The  data  processing is  described  in  TMH10  and \citet{HRCAM}  and
briefly  recalled here.   We use  the standard  speckle interferometry
technique  based on  the calculation  of  the power  spectrum and  the
speckle auto-correlation  function (ACF) derived  from it.  Companions
are detected  as secondary peaks in  the ACF and/or as  fringes in the
power spectrum.  Parameters of the binary and triple stars (separation
$\rho$, position angle $\theta$,  and magnitude difference $\Delta m$)
are determined by modeling the observed power spectrum.  Additionally,
the  true quadrant is  found from  the shift-and-add  images, whenever
possible.

The pixel scale  and angular offset are determined  by observations of
several relatively wide calibration  binaries. Their motion is modeled
based  on previous  observations at  SOAR, with  individual  scale and
orientation  corrections  for  each  observing run.   The  models  are
adjusted    iteratively    (the     latest    adjustment    in    2019
November). Measurements  of wide calibrators  by {\it Gaia}  show very
small  systematic  errors \citep{SAM18}.   Typical  rms deviations  of
observations of calibrators from their  models are 0\fdg2 in angle and
1 to  3 mas in separation.  The position accuracy  strongly depends on
the target characteristics (larger errors  at large $\Delta m$ and for
faint pairs), as well as on the seeing and telescope vibration. 

\begin{figure}
\epsscale{1.1}
%\plotone{dmplot.ps}
\plotone{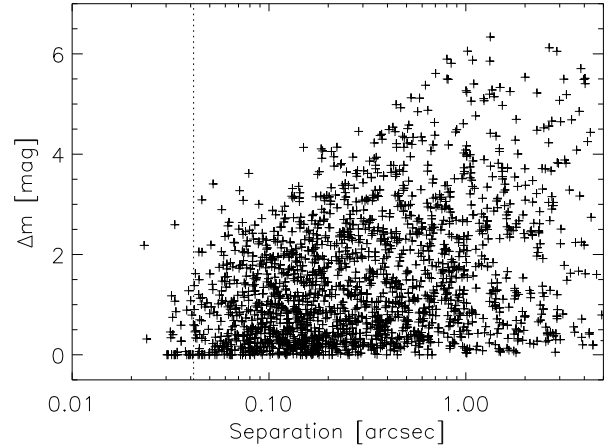}
\caption{Magnitude difference in the $I$ band vs. separation for pairs
  resolved in this filter.  The  vertical dotted line marks the formal
  diffraction limit of 41\,mas.
\label{fig:dm} }
\end{figure}

Figure~\ref{fig:dm} plots the  magnitude difference vs. separation for
pairs  resolved in  the $I$  filter  \citep[a similar  plot was  given
  in][]{SAM18}. The upper  envelope gives a clear idea  of the typical
contrast  limit  vs. separation.  Several  points  above the  envelope
correspond  to very  difficult tight  pairs with  a large  $\Delta I$;
these measures,  made at  or beyond the  limit of the  technique, have
large errors. Note points to  the left of the formal diffraction limit
(vertical dotted line, 41\,mas). 

%{\bf  Comment on super-diffraction  resolution, difficult  cases, wide
%  field, etc.}

%---------------------------------------------------------
\section{Results}
\label{sec:res}

\subsection{Data tables}

The  results  (measures of  resolved  pairs  and non-resolutions)  are
presented in  the same format  as in \citet{SAM18}.  The  long tables
are published electronically; here we describe their content.

\begin{deluxetable}{ l l  l l }
\tabletypesize{\scriptsize}
\tablewidth{0pt}
\tablecaption{Measurements of double stars at SOAR 
\label{tab:measures}}
\tablehead{
\colhead{Col.} &
\colhead{Label} &
\colhead{Format} &
\colhead{Description, units} 
}
\startdata
1 & WDS    & A10 & WDS code (J2000)  \\
2 & Discov.  & A16 & Discoverer code  \\
3 & Other  & A12 & Alternative name \\
4 & RA     & F8.4 & R.A. J2000 (deg) \\
5 & Dec    & F8.4 & Declination J2000 (deg) \\
6 & Epoch  & F9.4 & Julian year  (yr) \\
7 & Filt.  & A2 & Filter \\
8 & $N$    & I2 & Number of averaged cubes \\
9 & $\theta$ & F8.1 & Position angle (deg) \\
10 & $\rho \sigma_\theta$ & F5.1 & Tangential error (mas) \\
11 & $\rho$ & F8.4 & Separation (arcsec) \\
12 &  $\sigma_\rho$ & F5.1 & Radial error (mas) \\
13 &  $\Delta m$ & F7.1 & Magnitude difference (mag) \\
14 & Flag & A1 & Flag of magnitude difference\tablenotemark{a} \\
15 & (O$-$C)$_\theta$ & F8.1 & Residual in angle (deg) \\
16 & (O$-$C)$_\rho$ & F8.3 & Residual in separation (arcsec) \\
17  & Ref. & A8   & Orbit reference\tablenotemark{b} 
\enddata
\tablenotetext{a}{Flags: 
q -- the quadrant is determined; 
* -- $\Delta m$ and quadrant from average image; 
: -- noisy data. }
\tablenotetext{b}{References to VB6 are provided at
  \url{http://ad.usno.navy/mil/wds/orb6/wdsref.txt} }
\end{deluxetable}

Table~\ref{tab:measures}  lists 2555 measures  of 1972  resolved pairs
and  subsystems,  including  the   new  discoveries.   The  pairs  are
identified by  their WDS codes and discoverer  designations adopted in
the WDS catalog \citep{WDS}, as well as by alternative names in column
(3), mostly  from the {\it Hipparcos}  catalog. Equatorial coordinates
for the  epoch J2000 in  degrees are given  in columns (4) and  (5) to
facilitate matching with other catalogs and databases.  In the case of
resolved multiple systems, the position  measurements and their errors (columns
9--12) and  magnitude differences (column 13) refer  to the individual
pairings between  components, not to  their photo-centers.  As  in the
previous papers  of this series,  we list the internal  errors derived
from  the power  spectrum model  and from  the difference  between the
measures obtained from  two data cubes. The median  error is 0.4\,mas,
and 90\% of  errors are less than 1.8\,mas. The  real errors are usually
larger, especially for difficult pairs with substantial $\Delta m$
and/or    with    small    separations.    Residuals    from    orbits
(Section~\ref{sec:orbits})  and   from  the  models   of  calibrators,
typically between 1 and 5 mas rms, characterize the external errors of
the HRcam astrometry.

The  flags in column  (14) indicate  cases when  the true  quadrant is
determined (otherwise the position angle is measured modulo 180\degr),
when the  photometry of wide  pairs is derived from  the long-exposure
images (this  reduces the bias  caused by speckle  anisoplanatism) and
when the data are noisy  or the resolutions are tentative (see TMH10).
For binary stars with known  orbits, the residuals to the latest orbit
and its reference are provided in columns (15)--(17). This work is referenced as SOAR2019.

Non-resolutions are reported in Table~\ref{tab:single}. Its first
columns (1) to (8) have the same meaning and format as in
Table~\ref{tab:measures}. Column (9) gives the minimum resolvable
separation when pairs with $\Delta m < 1$ mag are detectable. It is
computed from the maximum spatial frequency of the useful signal in
the power spectrum and is normally close to the formal diffraction
limit $\lambda/D$. The following columns (10) and (11) provide the
indicative dynamic range, i.e. the maximum magnitude difference at
separations of 0\farcs15 and 1\arcsec, respectively. The last column
(12) marks noisy data by the flag ``:''. 

\begin{deluxetable}{ l l  l l }
\tabletypesize{\scriptsize}
\tablewidth{0pt}
\tablecaption{Unresolved stars 
\label{tab:single}}
\tablehead{
\colhead{Col.} &
\colhead{Label} &
\colhead{Format} &
\colhead{Description, units} 
}
\startdata
1 & WDS    & A10 & WDS code (J2000)  \\
2 & Discov.  & A16 & Discoverer code  \\
3 & Other  & A12 & Alternative name \\
4 & RA     & F8.4 & R.A. J2000 (deg) \\
5 & Dec    & F8.4 & Declination J2000 (deg) \\
6 & Epoch  & F9.4 & Julian year  (yr) \\
7 & Filt.  & A2 & Filter \\
8 & $N$    & I2 & Number of averaged cubes \\
9 & $\rho_{\rm min}$ & F7.3 & Angular resolution (arcsec)  \\
10&  $\Delta m$(0.15) & F7.2 & Max. $\Delta m$ at 0\farcs15 (mag) \\
11 &  $\Delta m$(1) & F7.2 & Max. $\Delta m$ at 1\arcsec (mag) \\
12 & Flag & A1 & : marks noisy data  
\enddata
\end{deluxetable}

Table~\ref{tab:measures}  contains  90 pairs  resolved  for the  first
time;  some of  those  were confirmed  in  subsequent observing  runs.
Additional first  resolutions belonging to  the projects led  by other
PIs will be  reported elsewhere (these pairs are  not published here),
while  new  pairs  discovered  in  Upper  Scorpius  are  published  by
\citet{Sco}.  In the following sub-section we discuss the new pairs.

%---------------------------------------------------------
\subsection{New pairs}
\label{sec:new}

\begin{figure*}
\epsscale{1.1}
%\plotone{Triples.eps}
\plotone{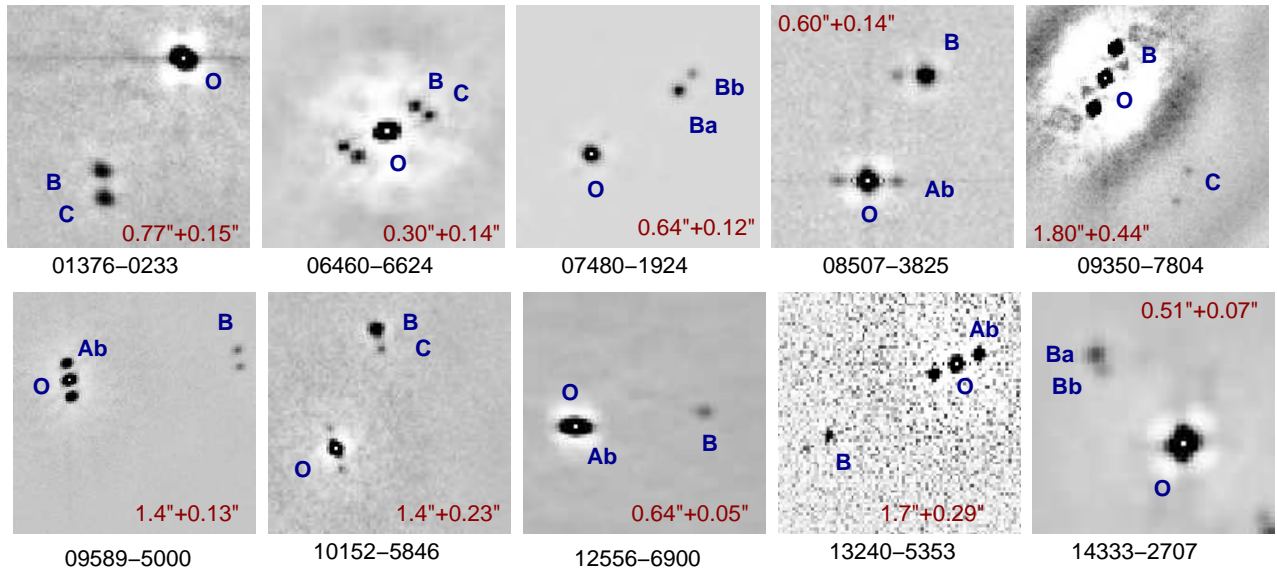}
\caption{Triple systems  discovered in 2019. Fragments  of the speckle
  ACFs  in arbitrary  negative intensity  scale are  shown.  The peaks
  corresponding to the components  are marked, the central peak (white
  dot at coordinate origin) is  labeled O. Separations of the wide and
  close pairs in arcseconds indicate the spatial scale. 
\label{fig:trip} }
\end{figure*}

\LongTables

\begin{deluxetable}{ l l c  c l  }
\tabletypesize{\scriptsize}
\tablewidth{0pt}
\tablecaption{New double stars
\label{tab:binaries}}
\tablehead{
\colhead{WDS} &
\colhead{Name} &
\colhead{$\rho$} &
\colhead{$\Delta m$} &
\colhead{Program\tablenotemark{a}} \\
 &     &    
 \colhead{(arcsec)} &
 \colhead{(mag)} & 
}
\startdata
00160$-$4816 &HIP 1276  &  0.19 &  0.4 & HIP \\
00271$-$3634 &HIP 2136  &  0.18 &  0.5 & HIP \\
00406$-$4831 &HIP 3186  &  0.04 &  0.8 & HIP \\ % closed 2019.86
00457$-$6752 &HIP 3579  &  0.08 &  2.6 & HIP \\
01043$-$5741 &SUB 1 Aa,Ab &  0.22 &  0.7 & HIP \\ % MSC, WD companion HIP 5018
01180$-$4809 &HIP 6075  &  0.32 &  1.5 & HIP \\
01376$-$0223 &RST4181 BC &  0.15 &  0.0 & WDS \\
02050$-$3748 &HIP 9713  &  0.34 &  3.1 & HIP \\
02065$+$0002 &HIP 9827  &  0.05 &  2.3 & REF \\
02143$-$4952 &HIP 10421 &  0.08 &  0.0 & HIP\tablenotemark{b} \\ % conf. 2020
02442$-$5234 &HIP 12775 &  1.33 &  4.4 & REF \\ % dy=5.5 missed in 2018
02466$-$3232 &HIP 12954 &  1.33 &  3.8 & HIP \\
02595$-$6415 &HIP 13935 &  0.51 &  4.6 & HIP \\
03274$-$4113 &HIP 16097 &  0.17 &  2.6 & HIP \\
03405$+$0508 &STF 430 Aa,Ab &  0.20 &  3.0 & REF \\ % AB optical. Giant.
03476$-$3625 &KPP2826 Aa,Ab &  0.45 &  3.2 & HIP \\ % +MSC
03566$-$3313 &HIP 18457 &  0.09 &  0.2 & HIP \\
04157$-$5631 &UC 1144 Aa,Ab &  0.19 &  2.3 & HIP \\ % triple
04249$-$3445 &DAM1313 Aa1,Aa2 &  0.06 &  1.8 & REF \\  % dy=2.1 phys. triple
05222$-$3218 &HIP 25085 &  0.07 &  2.6 & HIP \\ % TDS3161 (0.6'') is bogus.
06203$-$3004 &1 CMa &  0.04 &  0.0 & SB  \\ % 1 CMa P=675d
06225$-$6342 &HIP 30310 &  0.33 &  0.1 & HIP \\
06237$-$3319 &HIP 30410 &  0.07 &  0.1 & HIP\tablenotemark{b} \\ % Conf. 2019.9, retrograde.
06357$-$7006 &HJ 3885 Aa,Ab &  0.13 &  2.0 & WDS \\ % AB at 4''
06404$-$8223 &HIP 31931 &  1.66 &  2.5 & HIP \\
06460$-$6624 &HIP 32414 AB &  0.31 &  3.1 & HIP \\ %2019.9: triple, BC at 0.13 !
06460$-$6624 &HIP 32414 BC &  0.13 &  0.2 & HIP\tablenotemark{b} \\ %2019.9: triple, conf. ACF$-$CB.203.ps
07165$-$5513 &HIP 35203 &  0.17 &  1.6 & HIP\tablenotemark{b} \\  % Conf
07343$-$4517 &HIP 36818 &  0.09 &  3.0 & HIP \\ % optical ghost? needs conf.
07480$-$1924 &B 1077 Ba,Bb  &  0.12 & 0.7 & WDS \\ % AB 0.6'', AC 42'' ACF$-$CI.1247.ps
07548$-$6613 &HIP 38645 &  0.06 &  0.0 & HIP\tablenotemark{b} \\ %Conf, retrograde
08032$-$5401 &HIP 39391 &  0.96 &  2.7 & HIP\tablenotemark{b} \\ %Conf 2019.9.
08134$-$4534 &HIP 40269 &  0.04 &  0.0 & HIP\tablenotemark{b} \\ %Conf. 2019.9.
08170$-$3525 &HIP 40569 &  0.38 &  4.3 & HIP \\
08422$-$6852 &HIP 42709 &  0.67 &  1.8 & HIP \\
08507$-$3825 &JSP 308 Aa,Ab &  0.14 &  2.7 & WDS \\ % AB 0.6'' ACF$-$CI.1345.ps Re$-$do!
09012$+$0157 &CRC 57 Aa,Ab& 0.19 &  1.3 & MSC \\ % AB 3'', sb2, phys.  
09086$-$2960 &HIP 44868 &  1.68 &  5.3 & HIP \\
09350$-$7804 &KOH 85 AC  &  1.80  &  4.7 & WDS \\ % 2019.2: AC barely seen ACF$-$CB.260.ps
09448$-$3633 &HIP 47808 &  0.30 &  3.4 & HIP \\
09538$-$6719 &HIP 48528 &  0.36 &  3.2 & HIP\tablenotemark{b} \\ % Conf. 2019.9
09589$-$5000 &HIP 48928 AB & 1.37 &  1.9 & HIP\tablenotemark{b} \\ % conf, optical(DR2) ACF$-$CC.2422.ps
09589$-$5000 &HIP 48928 Aa,Ab &  0.06 & 0.0 & HIP\tablenotemark{b} \\ % conf,rapid motion. Check!
10152$-$5846 &HU 1596 BC &  0.23 &  2.4 & WDS \\ % AB 1.4'' ACF$-$CI.1450.ps
10212$-$1736 &HIP 50701 &  1.08 &  3.4 & HIP \\
10231$-$5032 &HIP 50861 &  1.06 &  3.6 & HIP\tablenotemark{b} \\ %Conf. 2019, almost fixed.
10343$-$7807 &HIP 51748 &  3.39 &  7.2: & HIP \\ %dI biased by truncation
10377$-$1103 &HIP 52023 &  0.44 &  3.6 & HIP\tablenotemark{b} \\ %Conf. 2019.9.
10560$-$0254 &HIP 53443 &  0.10 &  1.8 & HIP\tablenotemark{b} \\ % Conf. 2019.4, moves
11177$+$2722 &HIP 55170 &  0.07 & 1.4  & REF \\ % P*=8y
11415$-$7703 &HIP 57027 &  0.09 &  0.1 & REF \\ %P*=9y
11428$-$3549 &HIP 57129 &  1.26 &  2.5 & HIP \\
11515$-$2138 &HIP 57827 &  0.11 &  1.2 & HIP\tablenotemark{b} \\ %Conf. 2019.9, moves
11565$-$5046 &HIP 58226 &  0.75 &  3.1 & HIP\tablenotemark{b} \\ % conf 2020
12114$-$1647 &S 634 Aa,Ab & 0.023 & 0.7 & SB \\
12407$-$4803 &HIP 61868 &  0.06 &  1.7 & HIP\tablenotemark{b} \\ % conf, fast
12556$-$6900 &HDS1813Aa,Ab& 0.05 &  1.6 & HIP \\ % Aa,Ab UR 2019.5 but elong, proc?; AB 0.6''ACF$-$CB.449.ps
13103$-$3248 &HIP 64264 &  1.78 &  3.6 & HIP\tablenotemark{b} \\ % conf, var?
13240$-$5253 &HIP 65385 AB    &  1.70 &  2.0  & HIP \\ % new triple.AB is optical. ACF$-$CB.622.ps
13240$-$5253 &HIP 65385 Aa,Ab &  0.29 &  0.3 & HIP \\ 
13372$-$2337 &HIP 66433 &  0.28 &  3.7 & REF \\
14062$-$6543 &SKF 107 Aa,Ab &  0.98 &  3.6 & HIP \\ % AB 4.2'', optical,crowded
14079$-$3736 &HIP 69026 &  0.29 &  4.5  & HIP \\
14219$-$3609 &HIP 70214 &  1.09 &  3.9  & HIP \\
14333$-$2707 &HDS2056 Ba,Bb &  0.05 & 0.2 & HIP\tablenotemark{b} \\ % AB 0.5'' physical, Bab moves ACF$-$CE.393.ps
14333$-$3054 &HIP 71162 &  0.24 &  3.2  & HIP\tablenotemark{b} \\ % conf
14336$-$0956 &HIP 71188 &  1.28 &  3.3  & HIP \\
14347$-$3528 &HIP 71289 &  0.08 &  0.1  & REF \\
14386$-$0710 &HIP 71600 &  0.44 &  4.2  & HIP\tablenotemark{b} \\ % Conf. 2019.5.
14397$-$0957 &HIP 71685 &  1.48 &  4.1  & HIP \\
15031$-$4200 &B 1257 BC  &  0.15 &  0.8  & MSC \\ % AB 3.8'' ACF$-$CE.1241.ps
15031$-$4237 &WIS 279 Aa,Ab &  0.25 &  2.0  & HIP\tablenotemark{b} \\  % conf,triple
15107$-$4344 &CPO 415 Aa,Ab &  1.39 &  5.1  & REF \\ % optical
15594$-$3020 &HIP 78313 &  0.65 &  2.5  & HIP \\
16103$-$2209 &HIP 79244 &  0.03 &  0.6  & REF\tablenotemark{b} \\ % conf P*=3y! moves fast
16358$-$5345 &KPP3002 Aa,Ab &  1.17 &  2.6  & HIP\tablenotemark{b} \\ % Conf. 2019.6.
16486$-$3715 &HIP 82272 &  0.84 &  5.8  & REF \\
16520$-$3602 &HIP 82521 AB &  0.39 &  1.5  & HIP \\
16520$-$3602 &HIP 82521 Aa,Ab &  0.09 &  0.9  & HIP \\
17522$-$2440 &HIP 87453 &  2.65 &  6.1  & HIP \\
18431$+$0742 &LDS 1013 BC &  0.39 &  0.9 & MSC \\ % B=s2
18568$-$3002 &KPP4129 Aa,Ab &  0.13 &  2.0  & HIP\tablenotemark{b} \\ % conf
20363$-$1856 &TOK 339 Aa,Ab&  1.67 &  2.4  & HIP \\
20516$-$2927 &HIP 102963&  1.49 &  5.3  & HIP\tablenotemark{b} \\ % 2018.8 add, dy=6.9
22247$-$6537 &HIP 110630&  0.14 &  3.1  & REF\tablenotemark{b} \\ % conf
22374$-$4550 &SKF 384 Aa,Ab&  0.21 &  4.0  & HIP\tablenotemark{b} \\ %Conf. 2019.6 
22419$-$3155 &HIP 112064&  0.12 &  0.0  & HIP\tablenotemark{b} \\ %Conf. 2019.6
23232$-$5441 &HIP 115455&  1.62 &  2.6  & HIP\tablenotemark{b} \\ %Conf. 2019.6
23257$-$4537 &HIP 115648&  0.26 &  2.4  & HIP\tablenotemark{b} \\ %Conf. 2019.6
23308$-$4724 &HIP 116045&  0.23 &  2.4  & HIP\tablenotemark{b}  %Conf. 2019.6
\enddata 
\tablenotetext{a}{
HIP -- {\it Hipparcos} suspected binary;
MSC -- multiple system; 
REF -- reference star;
SB --  spectroscopic binary; 
WDS -- neglected pair.
}
\tablenotetext{b}{Confirmed.}
\end{deluxetable}

Table~\ref{tab:binaries} highlights  the 90 first-time  resolutions of
double  stars   or  new   subsystems  by  listing   their  approximate
separations at  discovery time,  magnitude differences (mostly  in the
$I$ band), and the corresponding observing programs. Full measurements
of these  pairs are found  in the  Table~\ref{tab:measures}.  Most
  new  pairs  (66) are  {\it  Hipparcos}  suspected
binaries.  We also resolved serendipitously 12 reference stars.  Seven
subsystems  in  the  previously   known  WDS  pairs,  also  discovered
accidentally, have  the program code  WDS. New resolutions  are either
very tight  pairs or wider pairs  with a large  contrast; all ``easy''
pairs  with comparable  components and  large separations  were already
discovered, e.g. by {\it Hipparcos}.

As  in  previous papers  of  this  series,  we discovered  new  visual
multiple  systems  with  three  or  more  resolved  components.   This
information  will  be  ingested   into  the  current  version  of  the
multiple-star  catalog, MSC  \citep{MSC}.   In the  case of  HIP~48928
(J09589$-$5000)  and HIP~65385 (J13240$-$5253),  both inner  and outer
pairs are  new discoveries.  Figure~\ref{fig:trip}  presents fragments
of  the speckle ACFs  of some  multiple systems.   Three new  wide and
faint  companions  to previously  known  closer pairs  (J03302$-$7024,
J09350$-$7804,  and J09589$-$500)  are  optical, as  revealed by  {\it
  Gaia}  DR2 by  their  discrepant proper  motions and/or  parallaxes.
Similarly, the known 4\farcs2 pair SKF~107 (J14062$-$6543) is optical,
while the chance of the new 1\arcsec ~pair being physical is higher.

New tight  binaries are promising candidates  for orbit determination.
HIP~79244 (J16103$-$2209)  shows fast  motion  during one  year. The  bright
spectroscopic binary HIP~30122 (J06203$-$3004, 1~CMa) with a period of
675 days was observed on  request by J.~Docobo and securely resolved into
a 35-mas near-equal pair.  Its single-lined spectroscopic orbit should
be upgraded to a double-lined  one to allow measurement of the orbital
parallax.

%-------------------------------------------------------------
\subsection{New and updated orbits}
\label{sec:orbits}

\begin{figure*}
\epsscale{0.9}
%\plotone{Cases.eps}
\plotone{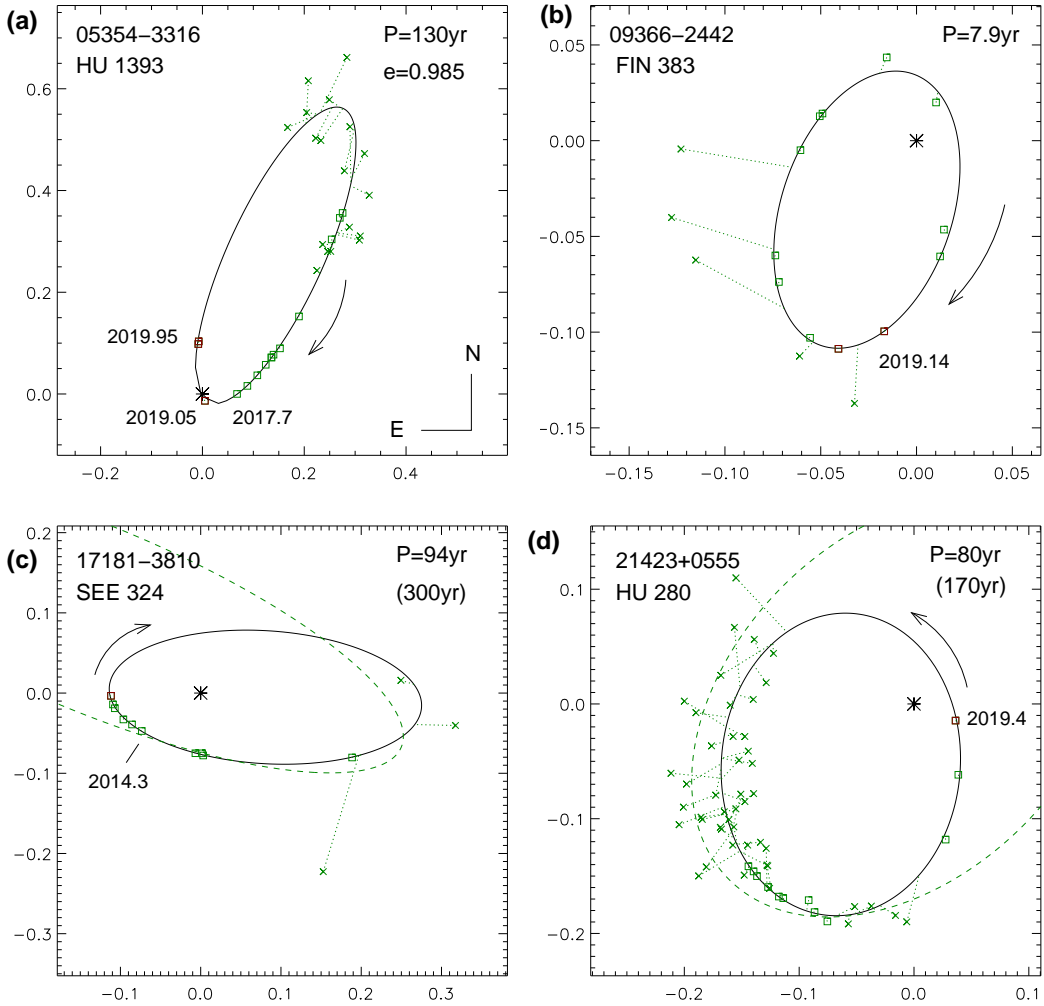}
\caption{Orbits of  ``classical'' visual binaries observed  at SOAR in
  2019.   In  this  and  following  Figures,  the  axis  scale  is  in
  arcseconds,  the  primary component  (asterisk)  is  located at  the
  coordinate     origin.      Green     squares    denote     accurate
  speckle-interferometric   measures   of   the  secondary   component
  (the 2019 measures are highlighted in red), crosses denote
  less accurate micrometer measures. The measures are connected to the
  orbit (ellipse) by short dotted  lines. In the two lower panels, the
  dashed lines are previous orbits.
\label{fig:cases} }
\end{figure*}

\begin{figure*}
\epsscale{0.9}
%\plotone{Mdwarfs.eps}
\plotone{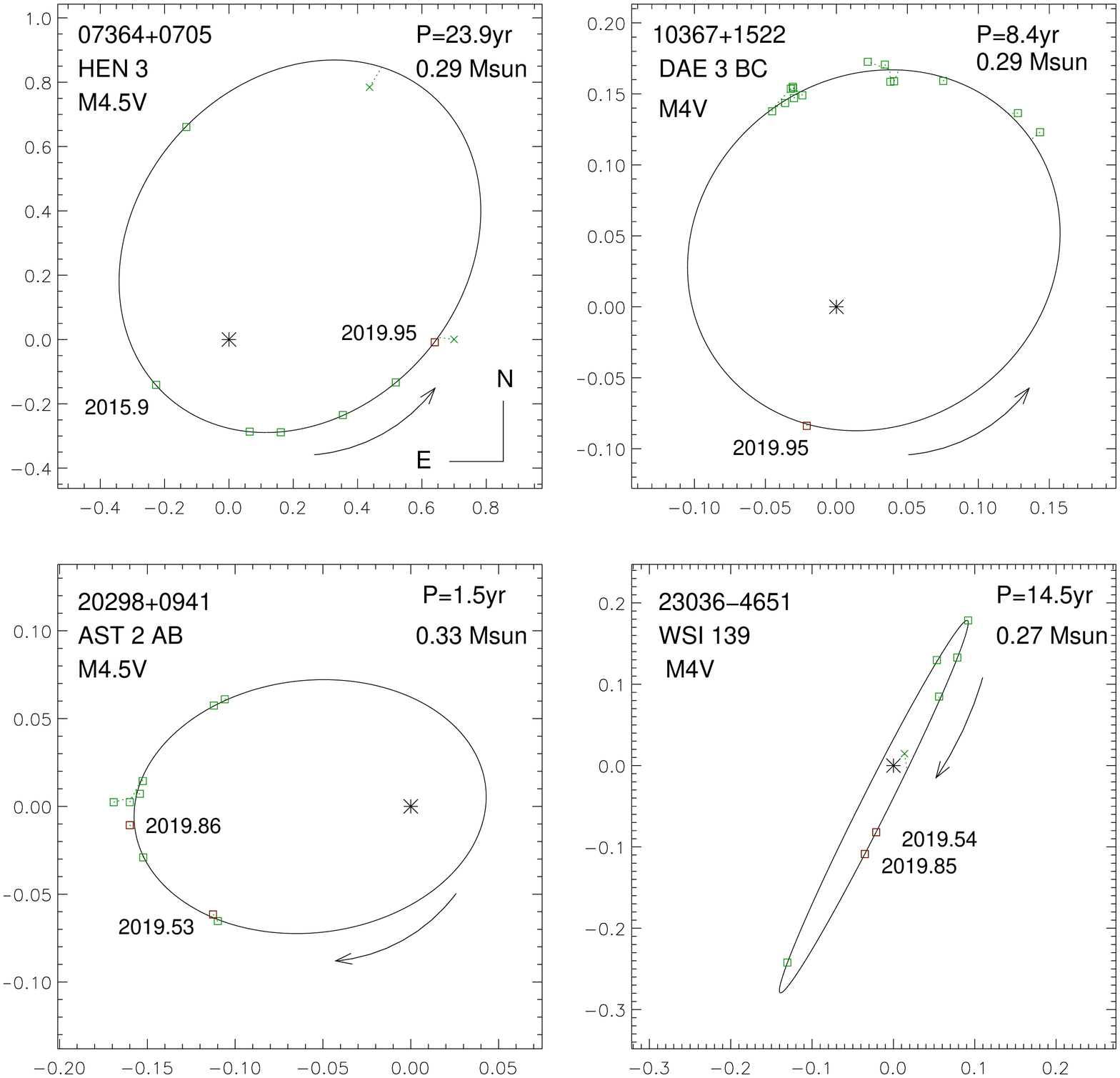}
\caption{New orbits of four M4V dwarf binaries. The mass sum computed
  from the orbital elements and parallax is indicated. 
\label{fig:dwarfs} }
\end{figure*}

Speckle  measurements at  SOAR are  used  to compute  new and  improve
previously known orbits. The  well-known difficulties inherent to visual
binaries  (insufficient coverage,  inaccurate or  misleading measures)
resulted in  the large  number of poor-quality  orbits in  the general
orbit catalog,  VB6 \citep{VB6}.  The  situation is improving  as new,
substantially more accurate data  become available.  At the same time,
many  first-time orbits  just  computed from  recent observations  are
tentative (grades 4 or 5) and contribute to the pool of poor orbits in
the catalog as the older orbits  become better known. In some cases, the
lack of coverage leaves  long-period orbits poorly constrained despite
good  modern  measures.  Nevertheless,  even  tentative (and  possibly
wrong)  orbits are  useful  in several  ways:  as a  synthesis of  all
existing  data,  for  predicting  binary  positions and  planning  future
measurements, etc.

In Table~\ref{tab:vborb}, orbital elements  and their errors are given
for  93 pairs  observed  in  2019.  Formal  grades  and references  to
previous orbits are given in the last columns (SOAR2019 for the orbits
computed here).   Asterisks mark  orbits where radial  velocities from
the  literature   are  used  jointly  with   position  measures.   The
complementary   Table~\ref{tab:vborb2}    lists 34  provisional,   poorly
constrained orbits  of grades  4 and 5  without errors, which  in this
case  are  large or  even  misleading.   For  circular and/or  face-on
orbits, some  Campbell elements become  degenerate and they  are fixed
accordingly.   Some reliable orbits  of grades  3 or  better resulting
from  our work  were already  published in  the  Information Circulars
\citep[e.g.][]{Circ199}.  However, the Circulars do not provide errors
of the elements,  so we publish these orbits here  with errors to give
the  full result  (some  with  a slight  adjustment  using the  latest
measures).   Orbit  correction is  a  continuous  process.  Note  that
Table~\ref{tab:vborb}  also contains  25 orbits  of  excellent quality
(grades  1  and 2).   The  official  grades  consider several  factors
\citep[see][]{VB6} and are not  uniquely correlated with errors of the
elements. For  this reason, 16 low-grade orbits  with reasonably small
errors are kept in Table~\ref{tab:vborb}.

The orbital elements and their  errors were determined by the weighted
least-squares fit using the IDL program {\tt ORBIT} \citep{ORBIT}.  We
adopt weights proportional to $\sigma^{-2}$, where the errors $\sigma$
are assigned according to the measurement technique (e.g.  from 2 to 5
mas for speckle  interferometry with 4-m class telescopes,  10 mas for
{\it Hipparcos}, 50 mas or  larger for visual micrometer measures) and
corrected iteratively to reduce  the impact of outliers, if necessary.
In some cases the published  radial velocities (RVs) are used, leading
to the combined spectro-interferometric orbits. Mass sums are computed
as  a   sanity  check,  especially  helpful   for  poorly  constrained
preliminary  orbits. Two pairs  (J04303+1950 and  J05229$-$4219) were
unresolved  in  2019 despite  predictions  of  their previous  orbits.
These non-resolutions  are accounted for by the  new, corrected orbits.

Figure~\ref{fig:cases}  gives  four   noteworthy  examples  where  our
observations substantially  contributed to the knowledge  of orbits of
classical visual binaries.  Speckle monitoring of J05354$-$3316 at
SOAR (panel a) indicated diminishing separation, and in 2019 the pair passed
through  the periastron. In  2019.05 it  was unresolved;  however, the
elongated power spectrum was fitted with a  fixed $\Delta y = 1.3$ mag, yielding
a  separation  of 15  mas  (below  the  diffraction limit).  Two  good
measures  were  taken  after     periastron.   The  extremely  high
eccentricity of $e=0.985  \pm 0.002$ is well constrained  by our data.
If the periastron had been missed, we would have to wait for 130 years for
the  next  one.   The case  of  J09366$-$2442  in  panel (b)  shows  how
unreliable or even  misleading the visual measures of  close pairs can
be.  Including these  measures  in  the orbital  fit  would spoil  the
result, while the speckle data alone define the orbit quite well.  The
period of J17181$-$3810 (panel c) was revised from 300 to 94 yr.  This
pair was not  observed at all for 69 years,  from 1897 till 1966,
which is rather unusual. Based on three micrometer measures and only a
few  speckle data  available in  2014, the  previous 300  yr  orbit by
\citet{TMH15} was just a bold  guess! Now the elements are constrained
better, but the  long period means that this  orbit will remain poorly
known for decades.   Yet another case of radical  orbit revision (from
170 to 80  yr) is J21423+0555 (panel d).  This  pair now rapidly moves
through the  periastron and  in a few  more years its  eccentric orbit
will become definitive. 

To  illustrate the  impact of  SOAR observations  on the  knowledge of
stellar    masses,   we    plot   in    Figure~\ref{fig:dwarfs}   four
well-constrained orbits  of nearby M-type dwarfs,  continuing the work
of  \citet{Mason2018}. Full  analysis of  mass-luminosity  relation is
outside the  scope of this paper,  and the orbits  are highlighted here
only  to illustrate the  potential of  HRCam data  in this  area.  The
spectral types  in the Figure  are retrieved from Simbad.   All orbits
except the last one were  known previously, but with a lower accuracy.
For  example,  for J10367+1522  (DAE  3 BC)  \citet{Calissendorff2017}
found  an excess  of mass  which is  not confirmed  by our  new orbit,
constrained nicely by the 2019 measure.   This is a triple system of M
dwarfs  at 20\,pc  where the  outer orbit  is also  computed, although
it is  still  poorly  constrained.   In  contrast,  J20298+0941  (GJ  791.2),
previously listed  in the WDS  as a tight  visual triple AB+AC,  is in
fact a simple  binary, never resolved as triple  at SOAR or elsewhere;
another  companion D at  9\arcsec ~is  optical.  We  used the  RVs and
position  measures by \citet{Benedict2016}  to get  a well-constrained
combined  orbit.  The  RVs,  however, are  likely  biased because  the
resulting orbital  parallax of 93$\pm$3\,mas  contradicts the accurate
parallax measured by the  same authors (113.4$\pm$0.2\,mas), while the
mass ratio of 0.5 inferred  from the RV amplitudes contradicts $\Delta
J  =   2.3$  mag   measured  at  SOAR.    The  {\it   Gaia}  parallax,
133.8$\pm$1.4\,mas, is  obviously biased by the orbit.   The last pair
of M-dwarfs  in Figure~\ref{fig:dwarfs}, J23036$-$4651  (WSI 139), did
not  have previous  orbit determinations.   Nevertheless, its  14.5 yr
orbit  based  almost  entirely  on  the SOAR  data  is  already  quite
reliable; the rms residuals are 2\,mas.

\begin{figure}
\epsscale{1.1}
%\plotone{HIP23766.eps}
\plotone{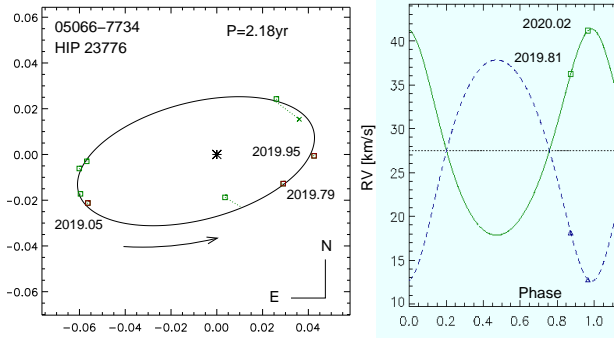}
\caption{Combined preliminary spectro-interferometric orbit of HIP 23776.
\label{fig:23776} }
\end{figure}

Figure~\ref{fig:23776}     illustrates    the     emerging    combined
spectro-interferometric         orbit         of         J05066$-$7736
(HIP~23776). Astrometric orbit  with a period of 1.9  yr based on {\it
  Hipparcos} was published by  \citet{Gln2006}. Our orbit with a period
of  2.18 yr  and  a semimajor  axis  of 53.1$\pm$0.9  mas is  strongly
constrained by  three measures taken  in 2019. All measures  come from
SOAR, and the rms residuals  to the orbit are 1.1 mas. High-resolution
spectroscopy (to be reported elsewhere) yields RV measurements of both
components. The preliminary  orbital parallax is 26.4$\pm$1.1\,mas and
the  masses are  0.89  and 0.83  solar.   The {\it  Gaia} parallax  is
28.68$\pm$0.18\,mas. Continued speckle interferometry and spectroscopy
of  this  pair  during  another   year  will  lead  to  accurate  mass
measurement.   The  spectral  type G8VFe-1.6CH-1.2  suggests  chemical
peculiarity, while  the fast proper  motion of 0\farcs45  yr$^{-1}$ is
typical  of  halo  or   thick  disk  stars.  \citet{GCS}  measured  the
metallicity  [Fe/H]  of  $-0.47$  and  determined the  mass  ratio  of
0.89$\pm$0.01;  our   preliminary  orbit  gives  the   mass  ratio  of
0.93$\pm$0.01.

% visual orbits
\LongTables
\begin{deluxetable*}{r l cccc ccc cc}    
\tabletypesize{\scriptsize}     
\tablecaption{Visual orbits
\label{tab:vborb}          }
\tablewidth{0pt}                                   
\tablehead{                                                                     
\colhead{WDS} & 
\colhead{Discov.} & 
\colhead{$P$} & 
\colhead{$T$} & 
\colhead{$e$} & 
\colhead{$a$} & 
\colhead{$\Omega$ } & 
\colhead{$\omega$ } & 
\colhead{$i$ } & 
\colhead{Grade }  &
\colhead{Ref.\tablenotemark{a}} \\
 \colhead{{\it HIP}} &
& 
\colhead{(yr)} &
\colhead{(yr)} & &
\colhead{(arcsec)} & 
\colhead{(deg)} & 
\colhead{(deg)} & 
\colhead{(deg)} &  & 
%\colhead{} &
%\colhead{} & 
}
\startdata
00121$-$5832 & RST4739 & 36.91 & 1989.95 & 0.597 & 0.236 & 55.0 & 278.2 & 147.4 & 3 & Tok2019h \\
{\it 975}    &    & $\pm$0.45 & $\pm$0.18 & $\pm$0.026 & $\pm$0.006 & $\pm$6.8 & $\pm$5.7 & $\pm$3.6&     &  \\
00277$-$1625 & YR 1 Aa,Ab & 13.35  & 2014.41  & 0.024 & 0.1208 & 167.8 & 270.3 & 69.6 & 4 & Tok2019h  \\
{\it  2190 }  &    & $\pm$0.15  & $\pm$0.54  & $\pm$0.019 & $\pm$0.0020 & $\pm$0.8 & $\pm$14.8 & $\pm$0.9&     &  \\
00462$-$2214 & RST4155 & 49.27  & 2003.49  & 0.305 & 0.1693 & 177.9 & 141.7 & 138.0 & 2 & Hei1984a \\
{\it  3606}   &    & $\pm$1.13  & $\pm$0.53  & $\pm$0.022 & $\pm$0.0044 & $\pm$4.3 & $\pm$2.5 & $\pm$2.3&     &  \\
00550$-$5315 & RST 23 & 140     & 1973.59  & 0.85  & 0.569  & 182.3 & 79.4 & 105.9 & 4 & SOAR2019 \\
 \ldots &    & $\pm$19     & $\pm$1.62  & $\pm$0.09  & $\pm$0.152  & $\pm$5.0 & $\pm$4.5 & $\pm$5.7&     &  \\
01308$-$5940 & TOK 183 & 6.586 & 2020.432 & 0.455 & 0.0671 & 352.3 & 42.4 & 93.5 & 3 & Tok2015c \\
{\it  7040}   &    & $\pm$0.059 & $\pm$0.079 & $\pm$0.017 & $\pm$0.0013 & $\pm$0.6 & $\pm$2.8 & $\pm$0.8&     &  \\
02038$-$0020 & TOK 38 Aa,Ab & 5.660 & 2005.426 & 0.367 & 0.0377 & 271.9 & 30.5 & 71.7 & 3 & SOAR2019* \\
{\it  9631}   &    & $\pm$0.006 & $\pm$0.020 & $\pm$0.006 & $\pm$0.0009 & $\pm$2.1 & $\pm$1.0 & $\pm$3.8&     &  \\
02290$-$1959 & RST2280 Aa,Ab & 31.41  & 2020.69  & 0.686 & 0.557  & 185.2 & 38.7 & 154.1 & 3 & Tok2018i \\
{\it  11565}  &    & $\pm$0.24  & $\pm$0.07  & $\pm$0.014 & $\pm$0.012  & $\pm$8.4 & $\pm$8.8 & $\pm$3.2&     &  \\
02305$-$4342 & ELP 1 Aa,Ab & 14.42  & 2020.0   & 0.032 & 0.130  & 122.3 & 14.7 & 139.7 & 3 & Tok2018c \\
 \ldots &    & $\pm$1.36  & $\pm$5.2   & $\pm$0.017 & $\pm$0.008  & $\pm$5.5 & $\pm$133.6 & $\pm$4.1&     &  \\
02517$-$5234 & HU 1562 & 67.62  & 2019.94  & 0.950 & 0.2588 & 49.1 & 168.9 & 128.9 & 3 & Tok2019h \\
{\it  13341}  &    & $\pm$1.07  & $\pm$0.17  & $\pm$0.022 & $\pm$0.0036 & $\pm$3.5 & $\pm$6.8 & $\pm$11.5&     &  \\
03271$+$1845 & CHR 10 AB & 16.88  & 1994.4  & 0.040 & 0.0787 & 91.2 & 206.7 & 42.8 & 3 & Ole1999 \\
{\it  16077}  &    & $\pm$0.10  & $\pm$3.3   & $\pm$0.032 & $\pm$0.0066 & $\pm$5.6 & $\pm$66.1 & $\pm$6.9&     &  \\
03379$+$0538 & YSC 27 & 16.51  & 2013.025 & 0.222 & 0.1399 & 98.6 & 94.7 & 123.5 & 3 & Cve2017b \\
{\it  16390}  &    & $\pm$0.20  & $\pm$0.066 & $\pm$0.012 & $\pm$0.0015 & $\pm$1.1 & $\pm$1.7 & $\pm$0.7&     &  \\
03462$-$2423 & RST2321 & 96.26  & 2013.99  & 0.832 & 0.2150 & 11.5 & 155.8 & 43.9 & 3 & SOAR2019   \\
{\it  17600}  &    & $\pm$2.43  & $\pm$0.40  & $\pm$0.027 & $\pm$0.0034 & $\pm$9.6 & $\pm$10.9 & $\pm$4.6&     &  \\
04008$+$0505 & A 1937 & 46.08  & 2014.74  & 0.535 & 0.0974 & 31.9 & 0.4 & 40.8 & 2 & Tok2019h \\
{\it  18374}  &    & $\pm$1.04  & $\pm$0.12  & $\pm$0.008 & $\pm$0.0014 & $\pm$3.0 & $\pm$3.4 & $\pm$2.0&     &  \\
04063$+$1952 & BAG 4 & 15.89  & 2012.25  & 0.921 & 0.0959 & 133.6 & 266.4 & 114.0 & 3 & Bag2001 \\
 \ldots &    & $\pm$0.12  & $\pm$0.39  & $\pm$0.010 & $\pm$0.0025 & $\pm$8.6 & $\pm$2.7 & fixed&     &  \\
04070$-$1000 & HDS 521 AB & 21.04  & 2017.938 & 0.712 & 0.2302 & 220.4 & 77.4 & 121.5 & 2 & Tok2019h \\
{\it  19206}  &    & $\pm$0.04  & $\pm$0.025 & $\pm$0.004 & $\pm$0.0020 & $\pm$0.4 & $\pm$0.3 & $\pm$0.5&     &  \\
04142$-$4608 & RST2338 & 18.145 & 2002.715 & 0.631 & 0.1941 & 169.1 & 138.9 & 30.6 & 2 & Doc2016i\\
{\it  19758}  &    & $\pm$0.054 & $\pm$0.088 & $\pm$0.012 & $\pm$0.0021 & $\pm$5.8 & $\pm$6.5 & $\pm$1.4&     &  \\
04303$+$1950 & PAT 10 & 11.016 & 1986.577 & 0.754 & 0.1280 & 127.6 & 34.1 & 142.1 & 4 & SOAR2019* \\
{\it  21008}  &    & $\pm$0.020 & $\pm$0.040 & $\pm$0.007 & $\pm$0.0019 & $\pm$2.6 & $\pm$3.0 & $\pm$2.0&     &  \\
04318$-$2407 & RST2347 & 122.9   & 2010.87  & 0.523 & 0.1734 & 163.5 & 339.7 & 129.4 & 3 & Tok2019h \\
{\it  21133}  &    & $\pm$8.2   & $\pm$0.44  & $\pm$0.025 & $\pm$0.0065 & $\pm$4.5 & $\pm$5.8 & $\pm$4.6&     &  \\
04389$-$1207 & HDS 599 & 50.13  & 2004.18  & 0.817 & 0.3174 & 153.9 & 281.2 & 76.5 & 3 & Tok2019h \\
{\it  21644}  &    & $\pm$1.45  & $\pm$0.30  & $\pm$0.018 & $\pm$0.0155 & $\pm$1.0 & $\pm$1.0 & $\pm$0.9&     &  \\
04406$-$0912 & WOR 17 & 215.5   & 1994.0   & 0.813 & 1.919  & 11.7 & 36.7 & 32.3 & 5 & SOAR2019 \\
{\it  21765}  &    & $\pm$20.8   & $\pm$0.35  & $\pm$0.014 & $\pm$0.086  & $\pm$4.6 & $\pm$5.1 & $\pm$3.4&     &  \\
04539$-$2032 & HDS 633 & 11.834 & 2018.197 & 0.913 & 0.1429 & 167.6 & 157.9 & 56.0 & 3 & Tok2019d\\
{\it  22772}  &    & $\pm$0.107 & $\pm$0.072 & $\pm$0.013 & $\pm$0.0042 & $\pm$5.4 & $\pm$9.1 & $\pm$3.1&     &  \\
04553$-$0352 & RST4257 AB & 87.63  & 2011.40  & 0.566 & 0.2497 & 169.8 & 3.1 & 119.5 & 4 & SOAR2019 \\
 \ldots &    & $\pm$2.48  & $\pm$0.74  & $\pm$0.023 & $\pm$0.0142 & $\pm$2.9 & $\pm$4.7 & $\pm$3.2&     &  \\
05066$-$7734 & TOK 785   & 2.178 & 2020.088 & 0.181 & 0.0531 & 284.5 & 7.0 & 60.6 & 3 & SOAR2019* \\
{\it  23776}  &    & $\pm$0.018 & $\pm$0.034 & $\pm$0.014 & $\pm$0.0009 & $\pm$1.4 & $\pm$6.5 & $\pm$2.2&     &  \\
05069$-$2135 & DON 93 BC & 115.4   & 2002.63  & 0.339 & 1.209  & 89.6 & 257.6 & 150.0 & 4 & SOAR2019 \\
 \ldots &    & $\pm$3.1   & $\pm$0.45  & $\pm$0.030 & $\pm$0.016  & $\pm$4.7 & $\pm$5.6 & $\pm$3.5&     &  \\
05229$-$4219 & TOK 93 Aa,Ab & 5.99  & 2013.78  & 0.555 & 0.0625 & 238.2 & 93.4 & 64.4 & 3 & Tok2016e \\
{\it  15148}  &    & $\pm$0.12  & $\pm$0.12  & $\pm$0.036 & $\pm$0.0031 & $\pm$2.8 & $\pm$1.1 & $\pm$2.5&     &  \\
05245$-$0224 & MCA 18 Aa,Ab & 9.414 & 2011.046 & 0.361 & 0.0472 & 125.3 & 12.6 & 103.5 & 2 & Tok2015c \\
{\it  25281}  &    & $\pm$0.041 & $\pm$0.250 & $\pm$0.018 & $\pm$0.0016 & $\pm$1.3 & $\pm$9.3 & $\pm$1.8&     &  \\
05272+1758 & MCA 19 Aa,Ab & 15.914 &  2014.60  &   0.806 &   0.0757 &   251.9 &   317.1 & 109.4 &   2  &  Jte2018 \\
{\it  25499}      &     &$\pm$0.073     &$\pm$0.16  &$\pm$0.083 &$\pm$0.0036 &$\pm$1.7  &$\pm$5.6  &$\pm$7.3 & & \\
05354$-$3316 & HU 1393 & 130.0   & 2019.04  & 0.985 & 0.6590 & 205.9 & 73.7 & 113.5 & 3 & FRM2014a \\
{\it  26245}  &    & $\pm$4.0   & $\pm$0.030 & $\pm$0.002 & $\pm$0.0515 & $\pm$1.4 & $\pm$1.5 & $\pm$1.8&     &  \\
05365$+$2556 & CHR 203 & 13.56  & 1993.65  & 0.947 & 0.0848 & 135.9 & 50.0 & 160.0 & 3 & SOAR2019 \\
{\it  26322}  &    & $\pm$0.126 & $\pm$0.299 & $\pm$0.010 & $\pm$0.0181 & $\pm$9.4 & fixed & $\pm$53.7&     &  \\
06035+1941 & MCA 24 & 13.061 &  2006.52 &  0.808 & 0.0555 & 224.4  &  291.5 & 111.8  &  2 &  Msn1997a \\
 {\it 28691}  &        &$\pm$0.031  &$\pm$0.16 &$\pm$0.045 &$\pm$0.0022   &$\pm$3.3  &$\pm$9.2  &$\pm$7.8 & & \\
06138$-$2352 & JNN 50 Ba,Bb & 11.75  & 2020.36  & 0.74  & 0.263  & 18.8 & 295.4 & 120.7 & 4 & SOAR2019 \\
{\it  29568}  &    & $\pm$1.03  & $\pm$0.23  & $\pm$0.12  & $\pm$0.020  & $\pm$12.5 & $\pm$9.5 & $\pm$2.1&     &  \\
06146$-$0434 & CHR 164 Aa,Ab & 18.94  & 2017.28  & 0.762 & 0.0491 & 287.4 & 227.1 & 57.8 & 3 & Tok2019d\\
{\it  29629}  &    & $\pm$0.64  & $\pm$0.29  & $\pm$0.037 & $\pm$0.0046 & $\pm$7.0 & $\pm$12.1 & $\pm$6.1&     &  \\
06253$+$0130 & FIN 343 & 76.76  & 2020.49  & 0.360 & 0.1321 & 165.3 & 153.7 & 161.9 & 3 & Tok2019d\\
{\it  30547}  &    & $\pm$3.22  & $\pm$0.45  & $\pm$0.014 & $\pm$0.0023 & $\pm$21.8 & $\pm$26.9 & $\pm$4.6&     &  \\
06573$-$4929 & RST5253 AB & 50.66  & 2006.36  & 0.038 & 0.2170 & 148.8 & 87.5 & 72.2 & 3 & Hrt2012a \\
{\it  33455}  &    & $\pm$1.53  & $\pm$2.03  & $\pm$0.032 & $\pm$0.0041 & $\pm$0.6 & $\pm$15.2 & $\pm$0.7&     &  \\
07312$+$0210 & TOK 393 & 5.71  & 2015.83  & 0.094 & 0.0606 & 27.3 & 168.4 & 144.2 & 3 & Tok2017c \\
{\it  36557}  &    & $\pm$0.14  & $\pm$0.25  & $\pm$0.030 & $\pm$0.0026 & $\pm$7.8 & $\pm$19.7 & $\pm$5.5&     &  \\
07364$+$0705 & HEN3 & 23.93  & 2016.208 & 0.587 & 0.6402 & 85.7 & 57.8 & 14.4 & 4 & Tok2018e \\
 \ldots &    & $\pm$0.40  & $\pm$0.008 & $\pm$0.005 & $\pm$0.0051 & $\pm$6.2 & $\pm$6.4 & $\pm$1.6&     &  \\
07560+2342 & COU 929 &   45.09 &   1997.98 &  0.4746 &  0.2555  &  186.70 &   70.66 & 71.86 &  1 &  Hrt2009  \\
{\it  38755}      &   &$\pm$0.19   &$\pm$0.21   &$\pm$0.0034  &$\pm$0.0011  &$\pm$0.23  &$\pm$0.73 &$\pm$0.22 & & \\
08125$-$4616 & CHR 143 Aa,Ab & 33.08  & 2017.67  & 0.284 & 0.0723 & 172.7 & 253.5 & 69.9 & 3 & Tok2015c \\
{\it  40183}  &    & $\pm$0.46  & $\pm$0.18  & $\pm$0.015 & $\pm$0.0010 & $\pm$1.0 & $\pm$2.7 & $\pm$1.1&     &  \\
08158$-$1027 & RST3578 AB & 29.37  & 2017.413 & 0.470 & 0.2220 & 93.3 & 216.8 & 59.3 & 3 & Tok2019d\\
{\it  40465}  &    & $\pm$0.09  & $\pm$0.076 & $\pm$0.019 & $\pm$0.0043 & $\pm$1.1 & $\pm$2.4 & $\pm$1.7&     &  \\
08230$-$7102 & HDS1196 & 19.45  & 2021.85  & 0.63  & 0.0652 & 16.3 & 175.7 & 133.6 & 3 & SOAR2019 \\
{\it  41093}  &    & $\pm$1.72  & $\pm$1.33  & $\pm$0.25  & $\pm$0.0079 & $\pm$18.6 & $\pm$32.2 & $\pm$24.5&     &  \\
08255$-$4058 & RST3592 & 133.8   & 2009.420 & 0.512 & 0.1442 & 165.0 & 246.2 & 140.4 & 3 & SOAR2019 \\
 \ldots &    & $\pm$19.2   & $\pm$0.46  & $\pm$0.050 & $\pm$0.0179 & $\pm$10.8 & $\pm$11.8 & $\pm$11.1&     &  \\
09252$-$1258 & WSI 73 & 12.715 & 2018.157 & 0.627 & 0.1294 & 93.1 & 42.7 & 85.9 & 4 & Tok2017b \\
{\it  46191}  &    & $\pm$0.123 & $\pm$0.059 & $\pm$0.009 & $\pm$0.0025 & $\pm$0.5 & $\pm$2.0 & $\pm$0.4&     &  \\
09366$-$2442 & FIN 383 & 7.883 & 2017.428 & 0.638 & 0.0814 & 176.2 & 223.7 & 130.9 & 2 & Tok2018b \\
{\it  47159}  &    & $\pm$0.041 & $\pm$0.037 & $\pm$0.011 & $\pm$0.0020 & $\pm$2.9 & $\pm$3.8 & $\pm$2.3&     &  \\
09439$-$5738 & HDS1404 Aa,Ab & 21.55  & 2019.79  & 0.723 & 0.1345 & 165.2 & 4.2 & 59.0 & 3 & SOAR2019 \\
{\it  47736}  &    & $\pm$0.835 & $\pm$0.050 & $\pm$0.018 & $\pm$0.0108 & $\pm$3.5 & $\pm$5.6 & $\pm$2.9&     &  \\
09442$-$2746 & FIN 326 & 18.390 & 2020.924 & 0.506 & 0.1075 & 175.0 & 138.2 & 126.8 & 2 & Doc2013d \\
{\it  47758} &    & $\pm$0.086 & $\pm$0.088 & $\pm$0.015 & $\pm$0.0012 & $\pm$1.8 & $\pm$2.6 & $\pm$0.9&     &  \\
09474+1134 & MCA 34 AB & 15.203 &  2003.96  &   0.3129 &  0.11097&   203.09 &   22.3 &  76.26&   2  & Jte2018 \\
{\it  48029 }     &       &$\pm$0.019  &$\pm$0.11  &$\pm$0.0065  &$\pm$0.00084 &$\pm$0.50 &$\pm$2.8  &$\pm$ 0.49 &  & \\
10116$+$1321 & HU 874 & 17.974 & 2003.819 & 0.928 & 0.1528 & 111.6 & 317.9 & 82.4 & 2 & Hrt1996a \\
{\it  49929}  &    & $\pm$0.032 & $\pm$0.092 & $\pm$0.006 & $\pm$0.0055 & $\pm$0.6 & $\pm$2.6 & $\pm$0.8&     &  \\
10345$-$3721 & RST3706 & 62.6   & 2025.8   & 0.292 & 0.1517 & 40.8 & 324.1 & 103.9 & 3 & SOAR2019 \\
{\it  51760}  &    & $\pm$6.343 & $\pm$1.773 & $\pm$0.092 & $\pm$0.0183 & $\pm$2.1 & $\pm$17.3 & $\pm$2.3&     &  \\
10367$+$1522 & DAE 3 BC & 8.427 & 2011.326 & 0.358 & 0.1382 & 104.1 & 44.4 & 21.4 & 4 & Jnn2017b \\
 \ldots &    & $\pm$0.029 & $\pm$0.041 & $\pm$0.009 & $\pm$0.0030 & $\pm$10.5 & $\pm$10.2 & $\pm$3.8&     &  \\
10373$-$4814 & SEE 119 & 16.672 & 2019.819 & 0.771 & 0.3999 & 35.7 & 287.0 & 123.3 & 2 & Tok2019d \\
{\it  51986}  &    & $\pm$0.032 & $\pm$0.005 & $\pm$0.003 & $\pm$0.0021 & $\pm$0.4 & $\pm$0.3 & $\pm$0.3&     &  \\
10397$-$3755 & HDS1523 & 59.7   & 2018.40  & 0.698 & 1.014 & 8.0 & 55.5 & 132.5 & 3 & SOAR2019 \\
{\it  52190}  &    & $\pm$1.24 & $\pm$0.002 & $\pm$0.003 & $\pm$0.015 & $\pm$0.6 & $\pm$0.5 & $\pm$0.6&     &  \\
11053$-$2718 & FIN 47 AB & 7.612 & 2013.749 & 0.370 & 0.1346 & 45.1 & 156.5 & 94.0 & 2 & Jte2018 \\
{\it  54204}  &    & $\pm$0.007 & $\pm$0.088 & $\pm$0.013 & $\pm$0.0015 & $\pm$0.4 & $\pm$3.9 & $\pm$0.4&     &  \\
11235$+$0701 & BAG 24 Aa,Ab & 20.810 & 2014.412 & 0.300 & 0.2275 & 150.0 & 222.0 & 160.0 & 3 & SOAR2019 \\
{\it  55605}  &    & $\pm$0.074 & $\pm$0.209 & $\pm$0.018 & $\pm$0.0016 & $\pm$16.8 & $\pm$18.9 & fixed&     &  \\
11446$-$4925 & RST9004 AB & 35.43  & 2009.866 & 0.409 & 0.3074 & 139.0 & 350.1 & 27.5 & 2 & Wol2014 \\
{\it  57269}  &    & $\pm$0.13  & $\pm$0.029 & $\pm$0.003 & $\pm$0.0014 & $\pm$1.8 & $\pm$1.9 & $\pm$0.8&     &  \\
12463$-$6806 & R 207 AB & 188.0   & 1874.2   & 0.79  & 1.16   & 138.6 & 72.5 & 53.3 & 3 & FMR2012g \\
{\it  62322}  &    & $\pm$16.5   & $\pm$6.5   & $\pm$0.13  & $\pm$0.23   & $\pm$12.1 & $\pm$11.3 & $\pm$9.9&     &  \\
13190$-$2536 & HDS1866 & 22.72  & 2008.29  & 0.397 & 0.1435  & 18.3 & 225.8 & 137.8 & 3 & SOAR2019 \\
{\it  64970}  &    & $\pm$0.37 & $\pm$0.16 & $\pm$0.024 & $\pm$0.0032 & $\pm$5.0 & $\pm$5.4 & $\pm$2.6 &     &  \\
13317$-$0219 & HDS1895 & 3.239 & 2013.781 & 0.534 & 0.0968 & 309.2 & 358.8 & 19.6 & 1 & Hrt2012a* \\
{\it  659820} &    & $\pm$0.002 & $\pm$0.006 & $\pm$0.006 & $\pm$0.0007 & $\pm$2.7 & $\pm$3.0 & $\pm$2.3&     &  \\
13334$+$0919 & HDS1902 & 26.15  & 2008.55  & 0.54  & 0.131 & 124.3 & 309.0 & 27.7 & 3 & SOAR2019 \\
{\it  66132}  &    & $\pm$0.51 & $\pm$0.44 & $\pm$0.01 & $\pm$0.016 & $\pm$34.2 & $\pm$34.5 & $\pm$15.9&     &  \\
13396+1045 & BU 612 AB   & 22.532 &  2019.7346&   0.5422 &  0.19932 &   36.71 &  356.9 &  45.34 &  1&   Msn1999a \\
{\it  66640}   &        &$\pm$0.020   &$\pm$0.0023  &$\pm$0.0023 &$\pm$0.00050  &$\pm$0.61  &$\pm$1.1 &$\pm$0.37& & \\
13520$-$3137 & BU 343 AB & 254.8   & 1996.36  & 0.646 & 1.0709 & 187.2 & 242.9 & 135.4 & 3 & Tok2014a \\
{\it  67696}  &    & $\pm$4.8   & $\pm$0.09  & $\pm$0.005 & $\pm$0.0100 & $\pm$0.6 & $\pm$0.9 & $\pm$0.4&     &  \\
13598$-$0333 & HDS1962 & 9.764 & 2008.366 & 0.405 & 0.0785 & 34.6 & 235.3 & 57.0 & 2 & Tok2018i \\
{\it  68380}  &    & $\pm$0.078 & $\pm$0.075 & $\pm$0.014 & $\pm$0.0016 & $\pm$1.6 & $\pm$2.3 & $\pm$1.2&     &  \\
14025$-$2440 & B 263 AB & 205.9   & 2011.84  & 0.533 & 0.597 & 19.8 & 65.5 & 48.8 & 4 & Tok2015c \\
{\it  68587}  &    & $\pm$17.8   & $\pm$0.64  & $\pm$0.024 & $\pm$0.037 & $\pm$1.6 & $\pm$4.8 & $\pm$1.7&     &  \\
14035$+$1047 & GJ 538 & 9.876 & 2011.177 & 0.480 & 0.3156 & 74.0 & 183.0 & 96.9 & 3 & SOAR2019* \\
{\it  68682}  &    & $\pm$0.002 & $\pm$0.005 & $\pm$0.001 & $\pm$0.0006 & $\pm$0.1 & $\pm$0.2 & $\pm$0.2&     &  \\
14382$+$1402 & TOK 406 & 8.29  & 2016.044 & 0.373 & 0.0998 & 7.3 & 168.3 & 130.5 & 3 & Tok2018e \\
{\it  71572}  &    & $\pm$0.14  & $\pm$0.028 & $\pm$0.010 & $\pm$0.0013 & $\pm$1.4 & $\pm$2.0 & $\pm$1.0&     &  \\
14581$-$4852 & WSI 80 & 22.39  & 2020.52  & 0.791 & 0.4015 & 86.6 & 109.2 & 108.4 & 3 & Tok2015c \\
{\it  73241}  &    & $\pm$2.08  & $\pm$0.19  & $\pm$0.061 & $\pm$0.0108 & $\pm$3.6 & $\pm$1.1 & $\pm$0.5&     &  \\
16059$+$1041 & HDS2273 Aa,Ab & 33.00  & 1992.977 & 0.426 & 0.3512 & 160.6 & 234.6 & 35.1 & 3 & Tok2019h \\
{\it  78864}  &    & $\pm$0.13  & $\pm$0.055 & $\pm$0.004 & $\pm$0.0021 & $\pm$1.2 & $\pm$0.9 & $\pm$0.9&     &  \\
16534$-$2025 & WSI 86 & 14.91  & 2014.96 & 0.502 & 0.2697 & 13.1 & 26.7 & 132.0 & 4 & SOAR2019 \\
{\it  82621}  &    & $\pm$0.25  & $\pm$0.17  & $\pm$0.013 & $\pm$0.0059 & $\pm$3.0 & $\pm$5.9 & $\pm$2.3&     &  \\
17156$-$1018 & BU 957 & 91.1   & 2030.2   & 0.570 & 0.287 & 25.0 & 28.8 & 101.1 & 2 & Tok2015c \\
{\it  84423}  &    & $\pm$2.9   & $\pm$1.8   & $\pm$0.048 & $\pm$0.013 & $\pm$1.2 & $\pm$4.2 & $\pm$1.1&     &  \\
17181$-$3810 & SEE 324 & 94.2   & 2020.54  & 0.414 & 0.195 & 266.3 & 179.9 & 117.8 & 3 & Tok2015c \\
{\it  84634}  &    & $\pm$12.3 & $\pm$1.55 & $\pm$0.028 & $\pm$0.007 & $\pm$1.9 & $\pm$14.1 & $\pm$3.0&     &  \\
17240$-$0921 & RST3972 Aa,Ab & 15.262 & 2006.017 & 0.570 & 0.1465 & 69.2 & 16.8 & 25.5 & 2 & Sod1999 \\
{\it  85141}  &    & $\pm$0.018 & $\pm$0.038 & $\pm$0.004 & $\pm$0.0012 & $\pm$3.5 & $\pm$3.9 & $\pm$1.8&     &  \\
17415$-$5348 & HDS 2502 & 20.39  & 2018.044 & 0.594 & 0.1341 & 158.5 & 321.5 & 136.7 & 3 & Tok2017c \\
{\it  86569}  &    & $\pm$0.34  & $\pm$0.028 & $\pm$0.008 & $\pm$0.0010 & $\pm$1.6 & $\pm$2.1 & $\pm$1.3&     &  \\
17586$-$1306 & HU 190 & 162.7   & 2011.40  & 0.422 & 0.3700 & 15.4 & 0.0 & 180.0 & 4 & Tok2019h \\
{\it  88010}  &    & $\pm$12.3   & $\pm$0.66  & $\pm$0.027 & $\pm$0.0142 & $\pm$3.2 & fixed & fixed&     &  \\
18480$-$1009 & HDS2665 & 38.44  & 2023.47  & 0.411 & 0.4796 & 36.4 & 195.7 & 56.4 & 4 & Tok2016e \\
{\it  92250}  &    & $\pm$1.22  & $\pm$0.30  & $\pm$0.038 & $\pm$0.0058 & $\pm$2.4 & $\pm$5.5 & $\pm$0.7&     &  \\
20048$+$0109 & TOK 699 & 7.83  & 2018.578 & 0.241 & 0.1579 & 74.1 & 45.1 & 36.3 & 3 & Tok2018i \\
{\it  98878}  &    & fixed & $\pm$0.037 & $\pm$0.008 & $\pm$0.0030 & $\pm$3.8 & $\pm$3.6 & $\pm$2.0&     &  \\
20210$-$1447 & BLA 7 Aa,Ab & 3.762 & 2015.642 & 0.452 & 0.0487 & 43.6 & 124.2 & 75.1 & 2 & Tok2015c \\
{\it  100325} &    & $\pm$0.000 & $\pm$0.009 & $\pm$0.009 & $\pm$0.0010 & $\pm$1.3 & $\pm$1.0 & $\pm$2.0&     &  \\
20298$+$0941 & AST 2 AB & 1.4708& 2019.185 & 0.573 & 0.1118 & 287.7 & 20.4  & 148.5 & 3 & Tok2019h* \\
 \ldots &    & $\pm$0.0004& $\pm$0.009 & $\pm$0.010 & $\pm$0.0015 & $\pm$2.4 & $\pm$3.6  & $\pm$3.0&     &  \\
20462$+$1554 & WSI 110 Aa,Ab & 4.871 & 2003.255 & 0.254 & 0.0979 & 143.6 & 344.0 & 78.5 & 3 & SOAR2019* \\
{\it  102490} &    & $\pm$0.001 & $\pm$0.018 & $\pm$0.006 & $\pm$0.0015 & $\pm$1.0 & $\pm$1.5 & $\pm$1.5&     &  \\
21088$-$0426 & HDS3013 Aa,Ab & 25.019 & 2019.391 & 0.557 & 0.3138 & 153.1 & 99.3 & 134.9 & 3 & Tok2018e \\
{\it  104383} &    & $\pm$0.026 & $\pm$0.017 & $\pm$0.007 & $\pm$0.0019 & $\pm$1.5 & $\pm$1.1 & $\pm$0.7&     &  \\
21094$-$7310 & I 379 AB & 5.389 & 2017.555 & 0.690 & 0.1829 & 194.4 & 183.4 & 93.3 & 2 & SOAR2019 \\
{\it  104440} &    & $\pm$0.059 & $\pm$0.041 & $\pm$0.022 & $\pm$0.0040 & $\pm$0.5 & $\pm$3.9 & $\pm$1.5&     &  \\
21130$-$1133 & VOU 24 AB & 128.5   & 2020.03  & 0.321 & 0.264  & 65.3 & 250.0 & 161.9 & 4 & SOAR2019 \\
 \ldots &    & $\pm$8.1   & $\pm$1.93  & $\pm$0.012 & $\pm$0.017  & $\pm$10.2 & fixed & $\pm$12.4&     &  \\
21198$-$2621 & BU 271 AB & 189.1   & 1841.48  & 0.631 & 2.159  & 244.7 & 190.5 & 65.2 & 3 & Tok2019h \\
{\it  105312} &    & $\pm$4.5   & $\pm$4.06  & $\pm$0.021 & $\pm$0.013  & $\pm$1.3 & $\pm$3.8 & $\pm$0.4&     &  \\
21395$-$0003 & BU 1212 AB & 48.68  & 1972.09  & 0.867 & 0.4277 & 141.1 & 293.6 & 55.2 & 2 & Tok2019h \\
{\it  106942} &    & $\pm$0.34  & $\pm$0.28  & $\pm$0.009 & $\pm$0.0093 & $\pm$1.9 & $\pm$1.6 & $\pm$0.8&     &  \\
21423$+$0555 & HU 280 & 80.77  & 2020.46  & 0.721 & 0.1778 & 198.2 & 103.3 & 51.8 & 2 & USN2006a \\
{\it  107153} &    & $\pm$2.52  & $\pm$0.15  & $\pm$0.024 & $\pm$0.0074 & $\pm$4.9 & $\pm$2.9 & $\pm$2.2&     &  \\
21477$-$3054 & FIN 330 AB & 20.218 & 2007.26  & 0.427 & 0.1240 & 31.9 & 215.2 & 108.4 & 2 & Doc2013d \\
{\it  107608} &    & $\pm$0.134 & $\pm$0.16  & $\pm$0.026 & $\pm$0.0016 & $\pm$1.2 & $\pm$3.1 & $\pm$0.9&     &  \\
21522$+$0538 & JOD 23 AB & 9.08  & 2019.386 & 0.447 & 0.1406 & 163.7 & 0.0 & 0.0 & 4 & Tok2019c \\
{\it  107948} &    & $\pm$0.11  & $\pm$0.024 & $\pm$0.008 & $\pm$0.0013 & $\pm$1.2 & fixed & fixed&     &  \\
22056$-$5858 & B 548 & 77.03  & 2019.8   & 0.186 & 0.265  & 210.9 & 54.0 & 73.0 & 3 & SOAR2019 \\
{\it  109060} &    & $\pm$1.45  & $\pm$3.9   & $\pm$0.048 & $\pm$0.006  & $\pm$1.7 & $\pm$23.1 & $\pm$1.3&     &  \\
22220$-$3431 & B 557 Aa,Ab & 107.97  & 2020.025 & 0.893 & 0.3559 & 190.0 & 72.8 & 112.6 & 3 & Tok2019h \\
{\it  110419} &    & $\pm$4.79  & $\pm$0.133 & $\pm$0.019 & $\pm$0.0290 & $\pm$1.9 & $\pm$1.7 & $\pm$2.4&     &  \\
22342$-$1841 & HU 389 & 186.0   & 2006.32  & 0.417 & 0.2995 & 122.5 & 158.9 & 50.6 & 3 & Tok2019h \\
{\it  111406} &    & $\pm$12.2   & $\pm$1.24  & $\pm$0.024 & $\pm$0.0085 & $\pm$2.6 & $\pm$2.8 & $\pm$1.7&     &  \\
22474$+$1749 & WSI 93 & 25.45  & 2017.372 & 0.634 & 0.2391 & 1.6 & 88.8 & 34.2 & 3 & Tok2017c \\
{\it  112506} &    & $\pm$1.27  & $\pm$0.030 & $\pm$0.014 & $\pm$0.0083 & $\pm$4.6 & $\pm$5.5 & $\pm$1.8&     &  \\
22479$-$5705 & B 2059 & 43.17  & 2007.00  & 0.70  & 0.125  & 71.9 & 176.0 & 102.1 & 3 & SOAR2019 \\
{\it  112561} &    & $\pm$2.82 & $\pm$2.70 & fixed & $\pm$0.007 & $\pm$4.4 & $\pm$24.7 & $\pm$3.7&     &  \\
22532$-$3750 & HDS3250 Aa,Ab & 10.62  & 2011.77  & 0.28  & 0.1395 & 226.0 & 100.1 & 56.7 & 4 & SOAR2019 \\
{\it  113010} &    & $\pm$1.51  & $\pm$0.17  & $\pm$0.13  & $\pm$0.0053 & $\pm$4.7 & $\pm$12.4 & $\pm$5.8&     &  \\
23036$-$4651 & WSI 139 & 14.50  & 2015.48  & 0.230 & 0.2572 & 153.3 & 162.4 & 93.8 & 3 & SOAR2019 \\
 \ldots &    & $\pm$2.15  & $\pm$0.44  & $\pm$0.059 & $\pm$0.0257 & $\pm$0.8 & $\pm$14.7 & $\pm$0.5&     &  \\
23210$+$1715 & WSI 11 & 7.95  & 2020.44  & 0.372 & 0.0852 & 163.7 & 22.9 & 28.7 & 3 & Tok2017c \\
{\it  115288} &    & $\pm$0.10  & $\pm$0.18  & $\pm$0.043 & $\pm$0.0042 & $\pm$16.8 & $\pm$18.1 & $\pm$10.1&     &  
\enddata 
\tablenotetext{a}{References to VB6 are provided at
  \url{http://ad.usno.navy/mil/wds/orb6/wdsref.txt} }
\end{deluxetable*}

% visual orbits
\LongTables
\begin{deluxetable*}{l l cccc ccc cc}    
\tabletypesize{\scriptsize}     
\tablecaption{Preliminary visual orbits
\label{tab:vborb2}          }
\tablewidth{0pt}                                   
\tablehead{                                                                     
\colhead{WDS} & 
\colhead{Discov.} & 
\colhead{$P$} & 
\colhead{$T$} & 
\colhead{$e$} & 
\colhead{$a$} & 
\colhead{$\Omega$ } & 
\colhead{$\omega$ } & 
\colhead{$i$ } & 
\colhead{Grade }  &
\colhead{Ref.\tablenotemark{a}} \\
 & & 
\colhead{(yr)} &
\colhead{(yr)} & &
\colhead{(arcsec)} & 
\colhead{(deg)} & 
\colhead{(deg)} & 
\colhead{(deg)} &  & 
%\colhead{} &
%\colhead{} & 
}
\startdata
00325$-$6800 & DON 7 & 150     & 2037.8   & 0.186 & 0.608  & 175.0 & 337.0 & 180.0 & 4 & SOAR2019 \\
00345$-$0433 & D 2 AB & 777     & 2008.5   & 0.79  & 1.111  & 83.1 & 270.4 & 77.6 & 4 & Hrt2010a \\
02128$-$0224 & TOK 39 Aa,Ab &0.2595  & 1989.5599& 0.689 & 0.0144 & 241.0 & 74.3 & 30.0 & 3 & SOAR2019* \\
03178$-$1407 & HU 432 & 73.67  & 1979.23  & 0.152 & 0.2017 & 45.4 & 223.2 & 40.4 & 4 & SOAR2019   \\
05190$-$2159 & RST2375 & 164.019 & 2015.515 & 0.326 & 0.3205 & 154.4 & 262.0 & 51.8 & 4 & Hrt2011d \\
05505$-$5246 & B 1493 & 113.0   & 2005.00  & 0.67  & 0.1625 & 247.9 & 170.2 & 140.5 & 4 & SOAR2019 \\
06515$+$0358 & A 1956 & 110.9   & 2002.6   & 0.437 & 0.3705 & 67.2 & 354.5 & 115.1 & 4 & SOAR2019 \\
08342$-$0957 & HDS1226 & 54.6   & 2003.13  & 0.386 & 0.203  & 202.6 & 113.6 & 119.5 & 4 & SOAR2019 \\
08582$+$1945 & LDS3836 & 134.4   & 2006.25  & 0.661 & 3.234  & 38.5 & 91.0 & 20.0 & 5 & SOAR2019 \\
09074$-$4357 & B 1646 AB & 208     & 2029.4   & 0.60  & 0.188  & 15.9 & 161.1 & 116.2 & 4 & SOAR2019 \\
09077$+$1040 & CHR 257 & 80.9   & 2030.06  & 0.55  & 0.201  & 106.2 & 170.7 & 79.2 & 3 & SOAR2019 \\
09370$-$2610 & WSI 127 & 31.98  & 2010.47  & 0.0   & 0.3907 & 284.0 & 0.0 & 0.0 & 5 & SOAR2019 \\
09393$-$1013 & RST3652 & 500     & 2002.254 & 0.70  & 1.355  & 277.7 & 96.1 & 113.7 & 5 & SOAR2019 \\
10038$-$0823 & HDS1454 & 32.74  & 2010.60  & 0.70  & 0.087  & 177.9 & 334.6 & 124.8 & 4 & SOAR2019 \\
10193$-$1232 & RST3688 & 180     & 2005.08  & 0.544 & 0.6852 & 142.1 & 278.5 & 80.9 & 4 & SOAR2019 \\
10476$-$1538 & TOK 714 Aa,Ab & 9.32  & 2015.59  & 0.103 & 0.0572 & 120.2 & 43.3 & 117.1 & 4 & SOAR2019 \\
11064$-$3545 & DAW 132 AB & 163.3   & 1983.62  & 0.97  & 1.199  & 121.8 & 177.4 & 141.2 & 5 & SOAR2019 \\
11342$+$1101 & YSC 43 Aa,Ab & 10.97  & 2005.44  & 0.522 & 0.0666 & 351.8 & 77.2 & 101.3 & 4 & SOAR2019 \\
11495$-$1636 & TOK 717 Aa,Ab & 24.0   & 2018.05  & 0.145 & 0.1984 & 292.5 & 108.2 & 102.6 & 4 & SOAR2019 \\
13327$+$2230 & HDS1898 & 30.0   & 2001.8  & 0.82  & 0.3056 & 251.0 & 94.1 & 74.0 & 3 & FMR2013d \\
13344$-$4224 & SEE 182 & 480     & 2017.290 & 0.983 & 1.073  & 28.1 & 288.0 & 55.0 & 5 & SOAR2019 \\
13344$-$5931 & TOK 403 & 18.44  & 2021.11  & 0.385 & 0.1441 & 77.2 & 265.5 & 117.7 & 4 & SOAR2019 \\
14038$-$6022 & VOU 31 AB & 182   0 & 2031.30  & 0.60  & 0.788  & 251.9 & 171.8 & 125.4 & 4 & SOAR2019 \\
15234$-$5919 & HJ 4757 & 600     & 2111.7   & 0.404 & 1.275  & 123.9 & 218.7 & 127.8 & 4 & Hrt2010a \\
16038$+$1406 & HDS2265 & 53.7   & 2021.17  & 0.815 & 0.3544 & 1.3 & 153.1 & 63.6 & 4 & Tok2018e \\
17012$-$1213 & HU 163 & 160     & 2009.0   & 0.73  & 0.1994 & 154.7 & 28.6 & 93.8 & 4 & SOAR2019 \\
19197$-$2836 & B 433 AB & 250     & 2027.51  & 0.458 & 0.441  & 219.7 & 97.7 & 57.2 & 4 & SOAR2019 \\
19563$-$3137 & TOK 698 & 13.0   & 2016.29  & 0.292 & 0.1063 & 59.8 & 44.0 & 90.8 & 4 & SOAR2019 \\
20521$+$0205 & A 2286 AB & 273.0   & 2015.07  & 0.687 & 0.311  & 83.3 & 226.8 & 60.6 & 4 & SOAR2019 \\
21477$-$1813 & CHR 223 & 73.829 & 2025.728 & 0.311 & 0.1713 & 110.9 & 275.3 & 109.9 & 4 & Tok2015c \\
21491$-$7206 & HEI 598 & 120     & 2016.44  & 0.349 & 1.565  & 42.8 & 67.8 & 51.8 & 4 & SOAR2019 \\
22266$-$1645 & SHJ 345 AB & 3500     & 2020.45  & 0.913 & 14.558  & 289.1 & 158.4 & 34.3 & 4 & Hle1994 \\
22378$-$5004 & HDS3214 & 82.7   & 2020.14  & 0.75  & 0.145  & 110.6 & 64.3 & 138.0 & 4 & Tok2019d \\
22441$+$0644 & TOK 703 & 5.02  & 2015.93  & 0.316 & 0.0585 & 168.7 & 166.5 & 134.3 & 4 & SOAR2019 
\enddata 
\tablenotetext{a}{References to VB6 are provided at
  \url{http://ad.usno.navy/mil/wds/orb6/wdsref.txt} }
\end{deluxetable*}

%-------------------------------------------------------------
%\subsection{}

%-------------------------------------------------------------
\section{Summary}
\label{sec:sum}

Continued  monitoring of  close  visual binaries  at  SOAR results  in
gradual  improvement of the  orbits, especially  for tight  and nearby
pairs  with  short  periods like  HIP~23776  (Figure~\ref{fig:23776}).
Good-quality  visual orbits  coupled to  precise parallaxes  from {\it
  Gaia} will vastly extend  our knowledge of stellar masses. Knowledge
of  visual orbits is  needed in  various astrophysical  contexts, for
example for binaries hosting exoplanets.

The SOAR speckle  program has resulted in the discovery  of many new close
binaries and  subsystems.  This  list is extended  here by the  90 new
pairs, although {\it Gaia} reveals  some wide and faint new companions
as  unrelated (optical).   During  2019, the  core  program on  visual
multiples  has  been  supplemented  by  various  binary  surveys  like
high-resolution  follow-up  of  TESS  exo-planet  candidates  and  the
multiplicity survey of Upper Scorpius association.

%---------------------------------------------------------
%\subsection{Comments on other pairs}
%\label{sec:other}

\acknowledgments 

We thank the SOAR operators for efficient support of this program, and
the SOAR  director J.~Elias for allocating some  technical time.  This
work is based in part  on observations carried out under CNTAC programs
 CN2019A-2 and CN2019B-13.

R.A.M.   and  E.C.   acknowledge  support from  the  Chilean Centro  de
Excelencia  en Astrof\'{i}sica y  Tecnolog\'{i}as Afines  (CATA) BASAL
AFB-170002, and FONDECYT/CONICYT grant \# 1190038.

%R.A.M. acknowledges  support from the  Chilean Centro de  Excelencia en
%Astrof\'{i}sica  y   Tecnolog\'{i}as  Afines  (CATA)   BASAL  AFB-170002,  and
%FONDECYT/CONICYT grant \# 1190038.

This work  used the  SIMBAD service operated  by Centre  des Donn\'ees
Stellaires  (Strasbourg, France),  bibliographic  references from  the
Astrophysics Data  System maintained  by SAO/NASA, and  the Washington
Double Star  Catalog maintained  at USNO.  This  work has made  use of
data   from   the   European   Space   Agency   (ESA)   mission   Gaia
(\url{https://www.cosmos.esa.int/gaia}  processed  by  the  Gaia  Data
Processing      and     Analysis      Consortium      (DPAC,     {\url
  https://www.cosmos.esa.int/web/gaia/dpac/consortium} Funding for the
DPAC  has been provided  by national  institutions, in  particular the
institutions participating in the Gaia Multilateral Agreement.

{\it Facilities:}  \facility{SOAR}.

%\clearpage
%\landscape

%\LongTables
%\input{orbtable1.tex}


\begin{thebibliography}{99}

\bibitem[Benedict et al.(2016)]{Benedict2016}
Benedict, G. F., Henry, T. J., Franz, O. G., et al. 2016, AJ, 152, 141 
%BENEDICT G.F., HENRY T.J., FRANZ O.G.
%2016AJ....152..141B 

\bibitem[Calissendorff et al.(2017)]{Calissendorff2017}
 Calissendorff, P., Janson, M., K\"ohler, R., et al. 2017, A\&A, 604,82
%    2017A&A...604A..82C

\bibitem [Gaia collaboration(2018)]  {Gaia}
Gaia Collaboration, Brown, A. G. A., Vallenari, A., Prusti, T., et
al. 2018, A\&A, 595, 2 (Vizier Catalog 	I/345/gaia2).
%GAIA COLLABORATION, BROWN A.G.A., VALLENARI A., PRUSTI T.
%2016A&A...595A...2G

\bibitem[Goldin \& Makarov(2006)]{Gln2006}
Goldin, A. \& Makarov, V. V. 2006, ApJS, 166, 341
%2006ApJS..166..341G

\bibitem[Hartkopf, Mason \& Worley (2001)]{VB6} 
Hartkopf, W. I., Mason, B. D. \& Worley, C. E. 2001, AJ, 122, 3472 
%(see the current version at 
%\url{http://www.usno.navy.mil/USNO/astrometry/optical-IR-prod/wds/orb6.html})

\bibitem[Hartkopf et al.(2012)]{Hrt2012a} 
Hartkopf, W. I., Tokovinin, A.  \& Mason, B. D.  2012, AJ, 143, 42

\bibitem[Horch et al.(2015)]{Horch2015}
Horch, E. P., van Belle, G. T., Davidson, J. W., Jr. et al. 2015, AJ, 150, 151
%2015AJ....150..151H
%; Ciastko, Lindsay A.; Everett, Mark E.; Bjorkman, Karen S.

\bibitem[Horch et al.(2017)]{Horch2017}
Horch, E. P., Casetti-Dinescu, D. I., Camarata, M. A., et al. 2017, AJ, 153, 212

\bibitem[Horch et al.(2019)]{Horch2019}
Horch, E. I., Tokovinin, A., Weiss, S. A., et al. 2019, AJ, 157, 56
%2019AJ....157...56H

\bibitem[Mason et al.(2001)]{WDS}
Mason, B. D., Wycoff, G. L., Hartkopf, W. I., et al.  2001, AJ, 122, 3466 (WDS)
%Douglass, G. G. \& Worley, C. E. 2001, AJ, 122, 3466 (WDS)

\bibitem[Mason et al.(2018)]{Mason2018}
Mason, B. D., Hartkopf, W. I., Miles, K. N., et al. 2018, AJ, 155, 215
%2018AJ....155..215M

\bibitem[Mendez et al.(2017)]{Mendez2017} 
Mendez, R.~A., Claveria, R.~M., Orchard, M.~E., \& Silva, J.~F.\ 2017, AJ, 154, 187 
%2017AJ....154..187M

\bibitem[Nordstr\"{o}m et al.(2004)]{GCS}
Nordstr\"{o}m, B., Mayor, M., Andersen, J., et al.\ 2004, \aap, 419, 989 

\bibitem[Perryman et al.(1997)]{HIP}
Perryman, M. A. C., Lindegren, L., Kovalevsky, J. ey al. 1997, A\&A, 323, L49 

\bibitem[Tokovinin(2012)]{Tok2012a} 
Tokovinin, A.  2012, AJ, 144, 56

\bibitem[Tokovinin(2016a)]{Tok2016c}
Tokovinin, A.  2016a, AJ, 152, 138

\bibitem[Tokovinin(2016b)]{ORBIT}
Tokovinin, A.  2016b, ORBIT: IDL software for visual, spectroscopic,
and combined orbits. Zenodo, doi:10.2581/zenodo.61119

%\bibitem[Tokovinin(2017)]{Tok2017c}
%Tokovinin, A. 2017, AJ, 154, 110

\bibitem[Tokovinin(2018a)]{HRCAM}
Tokovinin, A. 2018a, PASP,  130, 5002

\bibitem[Tokovinin(2018b)]{MSC}
Tokovinin, A. 2018b, ApJS,  235, 6

%\bibitem[Tokovinin(2018c)]{twins}
%Tokovinin, A. 2018c, AJ,  155, 160

\bibitem[Tokovinin(2019)]{Circ199}
Tokovinin, A. 2019, Inf. Circ., 199, 1
% 2018IAUDS.194....1T


\bibitem[Tokovinin \& Brice\~no(2020)]{Sco}
Tokovinin, A. \& Brice\~no, C.  2020, AJ, 159, 15


\bibitem[Tokovinin et al.(2010b)]{SAM09}
Tokovinin, A. Cantarutti, R., Tighe, R., et al. 2010b, PASP, 122,  1483 

\bibitem[Tokovinin et al.(2016b)]{SAM}
Tokovinin A., Cantarutti, R., Tighe R.,  et al. 2016b, PASP, 128, 125003

\bibitem[Tokovinin, Mason, \& Hartkopf(2010a)]{TMH10}
Tokovinin, A., Mason, B. D., \& Hartkopf, W. I. 2010a,  AJ, 139, 743 (TMH10)



\bibitem[Tokovinin et al.(2014)]{TMH14}
Tokovinin, A., Mason, B. D., \& Hartkopf, W. I.  2014, AJ, 147, 123 

\bibitem[Tokovinin et al.(2015)]{TMH15}
Tokovinin, A., Mason, B. D.,  Hartkopf, W. I., et al. 2015, AJ, 150, 50 

\bibitem[Tokovinin et al.(2016a)]{SAM15}
Tokovinin, A., Mason, B. D.,  Hartkopf, W. I., et al. 2016a, AJ, 152,
116 


\bibitem[Tokovinin et al.(2018)]{SAM17}
Tokovinin, A., Mason, B. D.,  Hartkopf, W. I., et al. 
 2018, AJ, 155, 235

\bibitem[Tokovinin et al.(2019)]{SAM18}
Tokovinin, A., Mason, B. D.,  Mendez, R. A., et al. 
 2019, AJ, 158, 148


\bibitem[Ziegler et al.(2020)]{TESS}
Ziegler, C., Tokovinin, A., Brice\~no, C., et al. 2020, AJ,  159, 19



% \bibitem[ ()]{}



\end{thebibliography}
\end{document}